\RequirePackage{silence}

\documentclass[reprint,aps,showpacs,pre]{revtex4-1}
\usepackage{graphicx} 
\usepackage{tabu}
\usepackage{placeins}
\usepackage{amsmath}
\usepackage{amsfonts,amssymb,wasysym} 
\usepackage{xr-hyper} 
\newcommand*\Laplace{\mathop{}\!\mathbin\bigtriangleup}
\pacs{89.75.Hc,05.45.-a,87.23.Cc}

\makeatletter
\def\blfootnote{\xdef\@thefnmark{}\@footnotetext}
\makeatother
\begin{document}

\title{Master stability functions reveal diffusion-driven pattern formation in networks}

\author{Andreas Brechtel$^{1a}$, Philipp Gramlich$^{1a}$, Daniel Ritterskamp$^{2a}$, Barbara Drossel$^{1b}$, Thilo Gross$^{2b}$} 
\affiliation{$^1$Institute of Condensed Matter Physics, Darmstadt University of Technology\\
$^2$ Department of Engineering Mathematics, Merchant Venturers School of Engineering, University of Bristol, Woodland Road, Bristol BS8 1UB, UK. 
 }
\date{\today}
\blfootnote{$^a$ These authors contributed equally.}
\blfootnote{$^b$ These authors contributed equally.}

\begin{abstract}
We study diffusion-driven pattern-formation in networks of networks, a class of multilayer systems, where different layers have the same topology, but different internal dynamics. Agents are assumed to disperse within a layer by undergoing random walks, while they can be created or destroyed by reactions between or within a layer. We show that the stability of homogeneous steady states can be analyzed with a master stability function approach that reveals a deep analogy between pattern formation in networks and pattern formation in continuous space.
For illustration we consider a generalized model of ecological meta-foodwebs. This fairly complex model describes the dispersal of many different species across a region consisting of a network of individual habitats while subject to realistic, nonlinear predator-prey interactions. In this example the method reveals the intricate dependence of the dynamics on the spatial structure. The ability of the proposed approach to deal with this fairly complex system highlights it as a promising tool for ecology and other applications.   
\end{abstract}
\maketitle 

\begin{figure*}[ht]
\includegraphics[width=\textwidth]{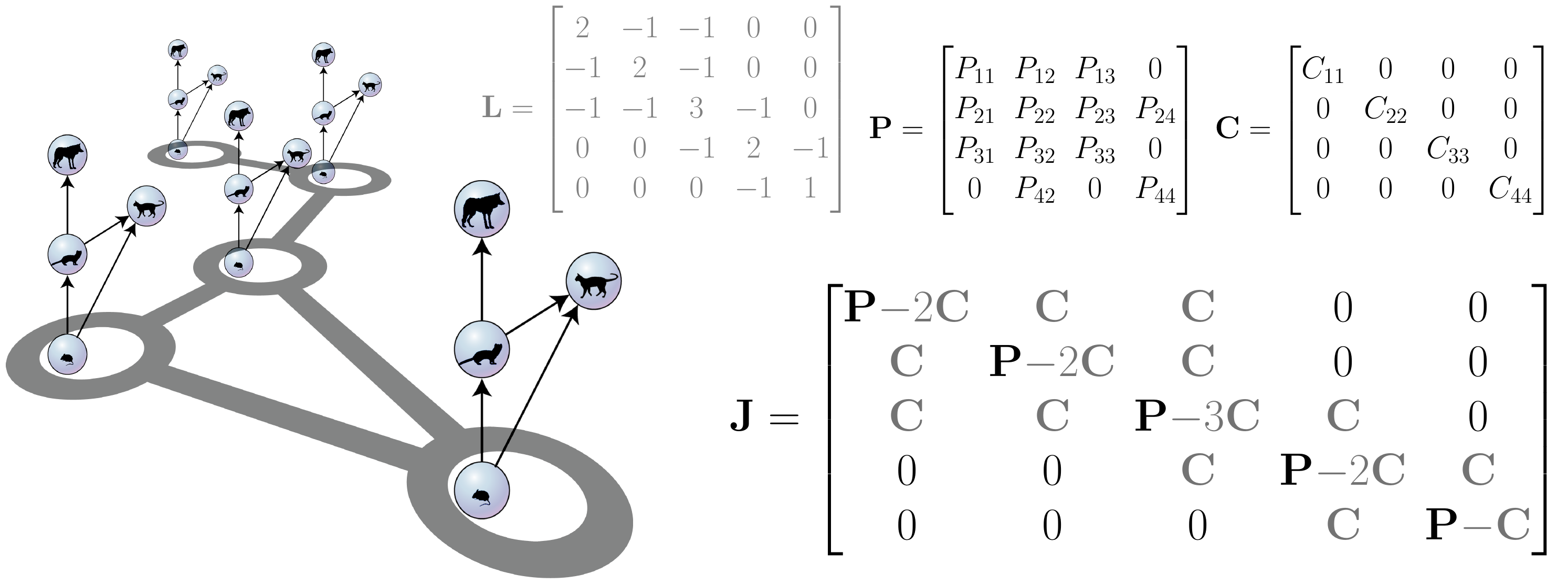}  
\caption{Example meta-foodweb with 4 species (blue bubbles and black arrows) on 5 habitats or patches (grey lines and circles). The stability of the system is described by the 20$\times$20 Jacobian matrix $\mathbf{J}$, which can be written as 5$\times$5 matrix of 4$\times$4 blocks. The blocks contain the intra-patch Jacobian matrix $\mathbf{P}$, describing the dynamics within one patch, and the coupling matrix $\mathbf{C}$, describing the dependence of migration rates on the population sizes, which is given here for a simple diffusion process. While the intra-patch Jacobian only appears on the diagonal of $\mathbf{J}$, the coupling blocks $\mathbf{C}$ occur in a pattern given by the Laplacian matrix $\mathbf{L}$ that encodes the structure of the patch network and is shown here for a coupling strength of 1.}  
\end{figure*}

\section{Introduction}
The study of complex networks has revealed many new interconnections between fields within the realm of complex systems, such as nonlinear dynamics and chaos, data analysis, cellular automata, graph theory, phase transitions and pattern formation. For these areas, networks provide an overarching mathematical framework that is conducive to a unification of complex system theory \cite{Newman2006}.
The present paper makes a contribution toward this goal by analyzing a general network formulation of diffusion-driven pattern formation. Our analysis points to an analogy between diffusive instabilities in continuous space and diffusive instabilities in networks. In both types of systems these instabilities can be described by structurally identical equations. Analyzing pattern formation in complex networks is therefore not more difficult than analyzing pattern formation in continuous space, but tends to produce richer behavior.   

Diffusion-driven instabilities in reaction-diffusion systems in continuous space were first discovered by Turing \cite{Turing1952} and later independently by Gierer and Meinhard \cite{Gierer1972}. For such systems it is well known that diffusion of one type of agent without reactions leads to a homogeneous distribution. Such homogeneous states also exist in reaction-diffusion systems where multiple species interact with themselves or each other while undergoing the diffusion process \cite{Turing1952}. In the continuous space reaction-diffusion system these homogeneous states are stationary but not necessarily stable to perturbations. When critical parameter values are crossed, their stability is lost in bifurcations, which mark the onset of pattern formation. These can be either Turing bifurcations, leading to stationary patterns, or wave instabilities, leading to travelling wave patterns \cite{Turing1952}.   

Nakao and Mikhailov \cite{Nakao2010} investigated an example of  a pattern forming instability on networks. In their systems agents of two species ``diffused'' by undergoing a random walk on a network. Reactions between the agents then led to a Turing-like diffusion-driven instability and subsequent pattern formation. 

In the present paper we take the idea of Nakao and Mikhailov further and formulate an analytical theory of diffusion-driven instabilities in networks.
We analyze systems that comprise a large number of different species subject to nonlinear interactions. There are thus two networks, the underlying geographical network on which the species diffuse, and a network of interactions between species. While the species differ in the nature of their interactions, they diffuse on the same geographical network, albeit potentially at different rates.  
We show that such network-on-network systems follow the same laws as the continuous space reaction-diffusion system, but can exhibit more complex behavior. 

Our derivation uses an approach that is formally equivalent to the Master-stability function technique that is widely used to study the stability of limit cycles in coupled oscillator systems \cite{Pecora1998,Arenas2008}. This work thus touches on two very active areas, the study of synchronization\cite{Pikovsky2001,Arenas2008} and multi-layer networks\cite{Parshani2010,Buldyrev2010,Bashan2013,DeDomenico2015}, which is the subject of several recent reviews \cite{Gao2011,Kivela2014,DeDomenico2014,Boccaletti2014,Cozzo2015,DeDomenico2016}. These two areas already have a wide interface due to many very recent papers, which studied synchronization on multilayer network\cite{Asllani2014,DeDomenico2014,Zhang2015,Kouvaris2015,DelGenio2016,Sevilla-Escoboza2016,Tang2016,Leyva2017}. 

The present work differs from the past papers in a number ways. Although some previous works investigated pattern formation in multilayer networks\cite{Asllani2014,Kouvaris2015}, these papers focused on inter-layer effects in networks where the layers have different topologies. By contrast we consider intra-layer effects, in a more restricted class of systems, which allows deeper analysis and thus reveals the analogy to the continuous case.

Previously the master stability function approach was widely used for the study of synchronization in multiplex networks\cite{DeDomenico2014,Zhang2015,DelGenio2016,Sevilla-Escoboza2016,Tang2016,Leyva2017}, but again the focus of these works was on inter-layer effects. Another difference to previous work is that we focus fundamentally on stationary states. Master stability function approaches have so far only been used in synchronization where they are applied to limit cycles. Although a synchronized state in a system of phase oscillators is mathematically a steady state one still thinks of it as a state in a system of coupled identical oscillators. Here we present a new derivation of the master stability function in the context of pattern formation and show that the application of the master-stability-function approaches to stationary states has great, presently untapped potential to lead to progress in applications.

We illustrate the potential of the approach by analyzing a complex ecological meta-foodweb model. If one were to reduce ecology to a single question, this question would probably focus on the origin and maintenance of biodiversity. In the words of  Hutchinson, ``Why are there so many kinds of animals?''\cite{Hutchinson1959}. Much current research in ecology is aimed at understanding how diversity is maintained and specifically how different species manage to coexist\cite{McCann2000}. Over the past 3 years two major approaches to this question have been converging. One of these is the study of ecological food webs, the networks of who-eats-who in ecology. Here, a major question is what properties stabilize the food web\cite{McCann2000,Gross2009,Kartascheff2010}. Theoretical insights in this area are still largely gained from simulations \cite{McCann2000}. However, the timescale separation between species on higher and lower trophic levels is a major obstacle that strongly limits the size of systems that can be simulated. 

The second avenue of investigation asks how spatial distribution affects the persistence of species. Early models considered only a single species in space, a so-called meta-populations\cite{Levins1969,Yeakel2014,Tromeur2016}. These models were subsequently extended to meta-communities, systems of competing similar species, e.g. different grasses in grassland plots. Only very recently predator-prey interactions  have been introduced to this line of ecological modeling \cite{Pillai2010,Pillai2011,Ristl2014,Gramlich2015,Barter2016,Mougi2016}. While the importance of such multilayer interactions for ecology has recently been emphasized \cite{Pilosof2015}, the considerable complexity of the resulting models has limited investigations to either a small number of species or a small number of spatial nodes. 

Using the approach proposed here, we are able to study complex meta-foodwebs, which we illustrate in a 30-species example system on a spatial network of arbitrary size. For this purpose we extend the so-called generalized food web model\cite{Gross2006,Gross2009,Gramlich2015}, to incorporate the dispersal of individuals across large networks of habitats or patches. The ecological model describes predator-prey interactions in continuous time. It uses bio\-lo\-gi\-c\-al\-ly realistic nonlinear interactions between species, and realistic scaling of dynamical timescales and diffusion rates with species body mass. For each such food web the proposed method yields a master stability function that reveals which spectral constraint the underlying geographical network must obey in order for the homogeneous steady state to be stable, and at what threshold parameter values pattern forming instabilities occur. These results remain valid for spatial networks of any size and provide the researcher with a principled approach to discussing what network features benefit stability and what hinders it. 

The paper starts by revisiting diffusion on networks and comparing it to diffusion in continuous space in Sec.~\ref{secDiffusion}. We then revisit diffusion-driven instabilities in continuous space in Sec.~\ref{secDDIcont} before deriving the master stability function approach for diffusion-driven instabilities on networks in Sec.~\ref{secDDI}. We propose the ecological model in Sec.~\ref{secModel} and present results from its analysis in Sec.~\ref{secResults}. Finally, we conclude with a discussion of the potential of the proposed approach for work in ecology and beyond in Sec.~\ref{secConclusions}.  

\section{Diffusion on networks \label{secDiffusion}}
Let us recapitulate some well-known properties of diffusion in continuous space that carry over to a properly defined diffusion process on networks:
\begin{itemize}
\item Let $X(x,t)$ be the concentration of agents or particles as a function of space and time. Then, given some initial configuration, the time evolution is given by     
\begin{equation}
\dot{X}=c\Laplace X
\label{eqEvolution}
\end{equation}
where $c$ is a diffusion constant and $\Laplace$ is the Laplace operator. 
\item This equation is solved by 
\begin{equation}
X(x,t) = \sum_n a_n{\rm e}^{c\kappa_n t} \boldsymbol{v_n} 
\label{eqDecomposition}
\end{equation}
where $\kappa_n$ and $\boldsymbol{v_n}$ are the eigenvalues and eigenfunctions of the Laplace operator on the spatial domain under consideration, and $a_n$ are expansion coefficients determined by the initial state. For instance, on a rectangular domain the eigenfunctions are trigonometric functions, and on a circular domain they are Bessel functions. In both cases, the eigenvalues $\kappa_n$ are wave numbers. 
\item On a connected domain, the Laplace operator is a negative semidefinite operator with a single zero eigenvalue, and the corresponding eigenfunction is the constant in space, such that 
\begin{equation}
\lim_{t\to\infty} X(x,t) = \mbox{const}\, .
\label{eqInfinity}
\end{equation}
\end{itemize}
Diffusion in networks has been studied for a long time \cite{Newman2006} and different types of diffusion processes on networks have been proposed. However perhaps the most intuitive process is the following: 

Consider a network of $N$ nodes described by an adjacency matrix $\bf A$ such that $A_{ij}=1$ if nodes $i$ and $j$ are connected and $A_{ij}=0$ otherwise. On this network let by $X_i(t)$ be the number of agents in node $i$. Agents undergo a random walk in continuous time, meaning that an agent has a constant probability density (per time) to transverse each link that is incident to its current node. Note that this means that agents leave nodes of higher degree more quickly, which is intuitive for instance if the agents are molecules diffusing between cavities in microfluidics and is also consistent with behavior observed in animals \cite{Hirt2017}.
In situations where coupling strengths, i.e.~link transversal probabilities, differ between or within networks, a weighted form of the adjacency matrix can be used where the nonzero elements $A_{ij}$  can be different from 1. Such situations arise for instance when diffusion through links is distance dependent.  

In the limit of large agent number the time evolution of the system can be written as 
\begin{equation}
\label{eqEvolutionNet}
\dot{\boldsymbol X}= -c {\bf L} {\boldsymbol X}
\end{equation}
where $\boldsymbol{X}=(X_1,\ldots,X_N)^{\rm T}$ and $c$ is a coupling constant and ${\bf L}$ is the Laplace matrix \cite{Newman2006}. The Laplace matrix is constructed by setting $L_{ii}=\sum_jA_{ij}$ and subtracting $\bf A$. For non-weighted networks where $A_{ij} \in \{0,1\}$, the diagonal elements of the Laplace matrix are identical to the degrees of the nodes (c.f.~Fig.~1 for illustration of the Laplacian). The Laplace matrix is aptly named as it can be interpreted as finite difference approximation to $-\Laplace$ on a network\cite{Merris1998}. 

Note that Eq.~(\ref{eqEvolutionNet}) is the analogous to Eq.~(\ref{eqEvolution}), except for an additional minus sign, which appears for historical reasons, as defining the Laplacian as a positive definite operator was thought to be advantageous in its original application. 

By decomposing $\boldsymbol X$ into eigenvectors of $\bf L$ we find the solution
\begin{equation}
\label{eqDecompositionNet}
\boldsymbol X(t) = \sum_n a_n {\rm e}^{-c\kappa_n t} \boldsymbol{v_n}
\end{equation}
where $a_n$ are again expansion coefficients determined by the initial state and $\kappa_n$ and $\boldsymbol{v_n}$ are the eigenvalues and eigenvectors of $\bf L$. 

Note that Eq.~(\ref{eqDecompositionNet}) is the network analogue of Eq.~(\ref{eqDecomposition}), except for the minus sign in the exponent. 

For the Laplacian matrix the row sum $\sum_j L_{ij}$ is zero in every row. Therefore, there is always an eigenvalue  $\kappa_1=0$, and the corresponding eigenvector is $\boldsymbol{v_1}=(1,1,\dots,1)$ \cite{Merris1998,Mohar1991,Agaev2005}. For connected networks this is the only zero eigenvalue of the Laplacian. Hence for $t\to\infty$ all terms of the Laplacian vanish except the $n=1$ term, and we are left with
\begin{equation}
\boldsymbol X(t\to \infty) = a_1 \boldsymbol{v_1}
\end{equation}
where $a_1$ ensures the correct normalization. This equation is the network analogue of Eq.~(\ref{eqInfinity}), as it implies that in the long run the system approaches a state where the concentration of agents is identical in each node. 

In this section we have revisited a well-known line of reasoning showing that a simple diffusion process on networks behaves analogously to a diffusion process in continuous space. In particular the Laplacian matrix of network science is the analogue of the Laplacian operator in continuous space and the system can be solved by a decomposition into the eigenmodes of this operator. In both cases this reveals that (on connected domains) only one eigenmode survives in the long-term dynamics which is constant in space. Thus the diffusion process approaches a state where the agents are uniformly distributed.


\section{Diffusion-driven instabilities in continuous space \label{secDDIcont}}
Let us now consider the case where multiple species $X_1,\ldots,X_M$ of agents diffuse over a network while undergoing reactions. 

In continuous space this is well studied\cite{Baurmann2007} and the following is known:
\begin{itemize}
\item The dynamics of a general reaction-diffusion system is given by equations of the form 
\begin{equation}
\dot{\boldsymbol X} = f(\boldsymbol{X}) + c\Laplace \boldsymbol X\, ,
\label{eqPDE}
\end{equation}
where $\boldsymbol X=(X_1(x),\ldots,X_M(x))$ is a vector of functions describing the distribution of the respective species in space, $f$ is a vector-valued function describing the local reactions, and $c$ is again a coupling constant. 
\item If the corresponding non-spatial system
\begin{equation}
\label{eqNonspatialCont}
\dot{\boldsymbol X} = f(\boldsymbol{X})
\end{equation} 
has a stationary state $\boldsymbol X^*$ then in the spatial system there is a corresponding homogeneous state in $\boldsymbol X^*(x)$, where the concentrations of agents are constant in space.  
\item The stability of the homogeneous steady states can be analyzed by linearizing the dynamics around the steady  state by setting $\boldsymbol X=\boldsymbol X^*+\delta$. This gives the equation
\begin{equation}
\dot\delta = {\bf J}\delta\, ,
\end{equation}
where $\bf J$ is the Jacobian matrix. The Jacobian matrix is a square matrix, whose linear dimension is the number of species $M$. We can compute it as 
\begin{equation}
J_{ab}= \left.\frac{\partial }{\partial X_b} \dot{X_a}\right|_* = \left.\frac{\partial }{\partial X_b} \left( f_a(\boldsymbol{X}) + c\Laplace X_a \right) \right|_*
\end{equation}
To avoid the spatial derivative we can again decompose the $\boldsymbol X$ into eigenfunctions $\boldsymbol{v_n}$ of the Laplace operator which yields a Jacobian matrix for every eigenmode $n$,   
\begin{equation}
J_{ab}^{(n)}= \left( \frac{\partial }{\partial X_b}  f_a(\boldsymbol{X})  \right)_* + c \kappa_n \delta_{ab} = P_{ab} + c \kappa_n \delta_{ab} 
\end{equation}
where $\kappa_n$ is the eigenvalue corresponding to $\boldsymbol{v_n}$, $\delta_{ab}$ is the Kronecker delta operator, and we absorbed the non-spatial derivatives into a new matrix $\bf P$. This matrix $\bf P$ is also the Jacobian of the corresponding non-spatial system, Eq.~(\ref{eqNonspatialCont}).

The system is stable with respect to a given eigen-perturbation if all eigenvalues of the corresponding Jacobian have negative real parts. The homogeneous steady state is thus stable if all the eigenmodes are stable, i.e.~if the eigenvalues of $\bf J$ are negative for all wave numbers $\kappa_n$.  We speak of a diffusion-driven instability when a change of parameters leads to the appearance of eigenvalues with positive real part for at least one non-zero wave number $\kappa_n$.
\end{itemize}


\section{Diffusion-driven instabilities in networks \label{secDDI}}
The beauty of the well-established method for the analysis of diffusions-driven instabilities, revisited above, is that the spatial (formally infinite-dimensional) system can be analyzed by considering the Jacobian matrix for the corresponding non-spatial system with some simple additional terms $c\kappa_n$ added to the diagonal elements. 

We now present a derivation that shows that in the network system a similar, equally elegant and equally simple approach is possible in which one can obtain the stability of spatial modes on the underlying geographical network by analyzing the Jacobian of the corresponding non-spatial system and then adding a minor modification that accounts for the nature of the eigenmode under consideration. 

As we now have to deal with two networks, the network of interactions between species and the underlying geographical network across which the agents diffuse we will avoid using the term `node' and instead use species to refer to a node of the species network and patch to refer to a node of the geographical network.  

Consider a reaction diffusion system on a network where $X_{ia}$ is the concentration of agents of species $a$ on patch $i$. The dynamics of the system is captured by the equation
\begin{equation}
\label{eqBasic}
\dot{X_{ia}} = \underbrace{f_a(\boldsymbol{X_i})}_{\mbox{ reactions}} - \underbrace{\sum_j c_a L_{ij}X_{ja}}_{\mbox{diffusion}},
\end{equation}
where $f_a$ is a function describing the impact of reactions on species $a$ depending on the local concentrations $\boldsymbol{X_i}=(X_{i1},\ldots,X_{iM})$, $c_a$ is the diffusion constant for species $a$, and $\bf L$ is again the Laplacian matrix. This equation, Eq.~(\ref{eqBasic}), is the network analogue of the continuous space equation Eq.~(\ref{eqPDE}). 

For comparison we also consider the corresponding non-spatial system
\begin{equation}
\label{eqNonspatial}
\dot{X_{a}} = f_a(\boldsymbol{X}),
\end{equation}
For any given stationary state $\boldsymbol{X^*}$ of Eq.~(\ref{eqNonspatial}) we can construct a homogeneous state of the network system Eq.~(\ref{eqBasic}) 
\begin{equation}
{X_{ia}}^* = {X_a}^* \quad \forall i
\end{equation}
We can quickly verify that the homogeneous states constructed in this way are stationary states of Eq.~(\ref{eqBasic}) because $X_{ia}^*$ satisfies $f_a(\boldsymbol{X_i})=0$ for all $i$ by construction. Furthermore due to the zero rowsum of the Laplacian it is true that ${\bf L} \boldsymbol{X_a}=0$ for all vectors $\boldsymbol{X_a}=(X_{1,a},\ldots,X_{N,a})$, where $X_{1,a}=\ldots=X_{N,a}$. 
 
Thus every stationary state of the non-spatial system corresponds to a stationary homogeneous state of the network system. This statement is the networks analogue of the statement from our second bullet point in the previous section.  

In the final, but most important step we now show that the stability of the homogeneous states can also be analyzed analogously to the continuous space system, i.e.~we can find the stability of network system by writing a Jacobian matrix that is identical to the Jacobian of the non-spacial system except for a minor modification. Analogously to the continuous case this modification depends on the ``wavenumber'' under consideration, i.e.~the eigenvalue of the respective Laplace operator. 

The network reaction-diffusion system is a large (i.e.~$NM$-dimensional) dynamical system. We can therefore analyze its stability by constructing the corresponding Jacobian matrix $\bf J$. To do this we must first bring the variables in linear order in a new vector 
\begin{equation}
\boldsymbol{Y}=(X_{1,1},\ldots,X_{1,M},X_{2,1},\ldots),
\end{equation}
i.e.~we order the variables such that the variables for all species in a given patch stay together. We can then compute the Jacobian matrix as 
\begin{equation}
J_{lm}=\left. \frac{\partial}{\partial Y_m} \dot{Y_l}\right|_*
\end{equation}
This matrix has the block structure illustrated in Fig.~1. We show this by computing the derivatives of the reaction and the diffusion terms separately (cf.~\ref{eqBasic}). 
The reaction rates at a given patch depend only on the concentrations in that patch. Therefore they vanish when differentiated with respect to concentrations in a different patch 
\begin{equation}
\frac{\partial}{\partial X_{jb}} f_a(\boldsymbol{X_i}) = 0 \quad \forall i \neq j.
\end{equation}
The derivative of a reaction term with respect to a variable in the same patch is identical to the corresponding derivative in the non-spatial system (Eq.~\ref{eqNonspatial}), 
\begin{equation}
\left. \frac{\partial}{\partial X_{ib}} f_a(\boldsymbol{X_i})\right|_* = P_{ab}\, .
\end{equation}
The contribution of the reaction terms to the Jacobian of the network system therefore has the form 
$$
\left(\begin{array}{c c c c} {\bf P} & 0 & 0 & \cdots \\
0 & {\bf P} & 0 & \cdots \\
0 & 0 & {\bf P} & \cdots \\
\vdots & \vdots & \vdots & \ddots\end{array}\right) \equiv {\bf I }\otimes {\bf P}\, ,
$$
where $\otimes$ denotes the Kronecker product. 
The identity matrix used here has a linear dimension equal to the number of patches $N$, while the 0 and ${\bf P}$ matrices have a linear dimension equal to the number of species $M$. 

Let us now consider the diffusion terms. Since we have assumed simple diffusion with a diffusion term that is linear in the concentration, we obtain directly
\begin{equation}
\left. \frac{\partial}{\partial X_{jb}} \sum_k c_a L_{ik}X_{ka}\right|_* = 
c_a L_{ij} \left. \frac{\partial}{\partial X_{jb}} X_{ja}\right|_* \equiv ({\bf L}\otimes{\bf C})_{ia,jb}
\end{equation}
with
$$
{\bf C}=\left(\begin{array}{c c c c} {c_1} & 0 & 0 & \cdots \\
0 & {c_2} & 0 & \cdots \\
0 & 0 & {c_3} & \cdots \\
\vdots & \vdots & \vdots & \ddots\end{array}\right).
$$
In the following, we will lift the restriction to simple diffusion, where the matrix $\bf C$ is diagonal. In certain applications the rates of diffusion can be a function of the concentrations of agents of the considered species and other species at the source. For example in an ecological context this allows to model animals that leave a patch more quickly in response to overcrowding when food is scarce or when predators are abundant in a patch. When we make again a linear approximation around the steady state, the diffusion term in 
Eq.~(\ref{eqBasic}) takes in this case the more general form
\begin{equation}
-\sum\limits_{j,b}c_{ab}L_{ij}X_{jb}\, .
\end{equation}
Now, the coefficients $c_{ab}$ become the elements of the matrix  $\bf C$.

Summarizing the above, we can write the Jacobian of the network system in the compact form
\begin{equation}
\mathbf{J}=\mathbf{I} \otimes \mathbf{P} - \mathbf{L} \otimes \mathbf{C}. \label{jacobian}
\end{equation}
The construction of the Jacobian from the different matrices is illustrated in Fig.~1.

Because the matrix has a block structure a similar structure exists in its eigenvectors. Consider vectors constructed as 
\begin{equation}
\boldsymbol{w} = \boldsymbol{v} \otimes \boldsymbol{q}  
\end{equation}
where $\boldsymbol{v}$ is a $N$-dimensional vector and $\boldsymbol{q}$ is an $M$-dimensional vector. 
Now, let $\boldsymbol{v}$ be an eigenvector of $\bf L$ with eigenvalue $\kappa$ such that 
\begin{equation}
{\bf L}\boldsymbol{v} = \kappa \boldsymbol{v}.
\end{equation}
Furthermore, let $\boldsymbol{q}$ be an eigenvector of $\mathbf P - \kappa \mathbf C$ with the eigenvalue $\lambda$.
Then $\boldsymbol{w}$ is an eigenvector of $\bf J$ to the eigenvalue $\lambda$, as the following calculation shows:
\begin{align}
\mathbf J \boldsymbol{w} 
	&= (\mathbf I \otimes \mathbf P - \mathbf L \otimes \mathbf C) \cdot (\boldsymbol{v} \otimes \boldsymbol{q}) \nonumber\\
	&= (\mathbf I \otimes \mathbf P) \cdot (\boldsymbol{v} \otimes \boldsymbol{q}) - (\mathbf L \otimes \mathbf C) \cdot (\boldsymbol{v} \otimes \boldsymbol{q}) \nonumber\\
	&= \mathbf I\boldsymbol{v} \otimes \mathbf P\boldsymbol{q} - \mathbf L \boldsymbol{v} \otimes \mathbf C \boldsymbol{q} \nonumber\\
    &= \boldsymbol{v} \otimes \mathbf P \boldsymbol{q} - \kappa \boldsymbol{v} \otimes \mathbf C\boldsymbol{q} \nonumber\\
	&= \boldsymbol{v} \otimes (\mathbf P - \kappa \mathbf C)\boldsymbol{q}\nonumber\\
    &=
 \boldsymbol{v} \otimes \lambda \boldsymbol{q} = \lambda (\boldsymbol{v} \otimes \boldsymbol{q}) =\lambda \boldsymbol{w}\,. \label{proof}
\end{align}

As all eigenvectors of $\bf J$ can be constructed in this way we can compute the complete spectrum of the network-level Jacobian as 
\begin{equation}
{\rm Ev}({\bf J}) = \bigcup_{n=1}^N{\rm Ev}({\bf P}-\kappa_n {\bf C}),
\label{qlambda}
\end{equation}
which enables us to analyze the stability of the network reaction-diffusion system by first computing the spectrum of the Laplacian matrix (the network analogue of wave numbers $\kappa$ in the spatial domain and then computing the eigenvalues of ${\bf P-\kappa {\bf C}}$, i.e.~by diagonalizing a matrix which is identical to the non-spatial Jacobian matrix $\bf P$ plus a minor modification $-\kappa {\bf C}$, which is specific to the respective eigenperturbation under consideration. In this sense Eq.~(\ref{qlambda}) establishes that diffusive instabilities in reaction-diffusion systems on networks can be computed using an approach that is analogous to the widely used approach to reaction diffusion systems in continuous space.  

In summary there exists a deep analogy between diffusion in continuous space and diffusion in networks. Above, we have shown that this analogy extends to diffusion-driven instabilities. In the network context the negative Laplacian matrix $- {\bf L}$ takes the role of the of the Laplace operator in continuous space. Correspondingly in the analysis the eigenvalues and eigenfunctions of the Laplace operator are replaced by the eigenvalues and eigenvectors of the Laplacian matrix. Analyzing diffusive instabilities is therefore not more complicated than analyzing those instabilities in continuous space. However, as networks tend to have more complex spectra more complex behavior can be expected.  

\begin{figure*}

\begin{tabular}{ m{4cm}  m{6cm}  m{6cm} }
\hline
      & \begin{center} \bfseries{Continuous Space} \end{center} &  \begin{center} \bfseries{Network} \end{center} \\\hline
Laplacian Operator  & \begin{center} Laplace Operator $\Laplace$ \end{center} & \begin{center} Laplace Matrix -$\bf L$ \end{center} \\
Eigenmodes  & \begin{center} $\Laplace v_n=\kappa_n v_n$  \end{center} & \begin{center} ${\bf L} \boldsymbol{v_n} = \kappa_n \boldsymbol{v_n}$ \end{center} \\
Reaction diffusion system  & \begin{center} $\dot{\boldsymbol X}=f(\boldsymbol{X})+{\bf C}\Laplace \boldsymbol{X}$  \end{center} & \begin{center}  $\dot{\boldsymbol X}_i=f(\boldsymbol{X}_i)-L_{ij}{\bf C} \boldsymbol{X}_j$ \end{center} \\
Diffusive Instability  & 
\begin{center} ${\rm Re}({\rm Ev}({\bf P}+\kappa_n{\bf C}))>0$  \end{center} & \begin{center} $ {\rm Re}({\rm Ev}({\bf P}-\kappa_n{\bf C}))>0 $ \end{center} \\\hline
\end{tabular}
\caption{Analogy between diffusion in continuous space and in networks. To emphasize the similarity we have written the network reaction-diffusion system, Eq.~(\ref{eqBasic}), in matrix form and also allowed a coupling matrix $\bf C$ in the instability condition for the continuous space systems. While our derivation considered the simpler case where $\bf C$ is proportional to the identity matrix, such matrices appear in the case of cross diffusion where diffusion of one species depends on the concentration of other species \cite{Baurmann2007}. While the concentrations $\boldsymbol{X}$ are position dependent in continuous space, the state of the system on a network is captured by a discrete set of variables $\boldsymbol{X}_i$, with $i$ being the node index.    See text for detailed derivations. } 
\end{figure*}


\section{Meta-Foodweb Model \label{secModel}}
In order to illustrate the powerfulness of this approach, we consider the example of a state-of-the-art meta-foodweb model from ecology, consisting of a set of identical local foodwebs, coupled in a spatial network (see Fig.~1). 

We build on the so-called generalized foodweb model. This fairly complex fully nonlinear model has been derived originally in \cite{Gross2006} and was used and validated in several recent ecological papers e.g.~\cite{Gross2009,Yeakel2014a}. While we refer the reader to \cite{Gross2006,Plitzko2012} for a full discussion of the model, let us revisit some of the design principles, before we extend the model to a spatial context.  

Consider a general system in which there is a population $X$ that is subject to a gain $G$ and a loss $L$. Without  knowing further details of the gain and loss processes we can model the system by 
\begin{equation}
\label{generalsmall}
\dot{X}=G(X)-L(X),
\end{equation}
where $G$ and $L$ are unspecified functions. Let us now assume that the system has a steady steady state $X^*$. This assumption is generally warranted in the sense that Eq.~(\ref{generalsmall}) defines a space of plausible models based on the available information. For every given $X^*$ we can find a specific model within this space that has $X^*$ as a steady state \cite{Kuehn2013}.  

We can now define a normalized variable $x=X/X^*$ and normalized functions $g(x)=G(X)/G(X^*)$, $l(x)=L(X)/L(X^*)$, which allow us to rewrite Eq.~(\ref{generalsmall}) as 
\begin{equation}
\dot{x}=\alpha(g(x)-l(x)),
\end{equation}
where $\alpha=G^*/X^*=L^*/X^*$. In the normalized system the stationary state under consideration is $x^*=1$ and its stability is determined by the Jacobian 
\begin{equation}
\label{smalljacobian}
{\bf P}=\alpha (\gamma - \mu), 
\end{equation}
where 
\begin{equation}
\gamma = \left.\frac{\partial}{\partial x}g(x)\right|_1 = \left. \frac{\partial }{\partial {\rm log}X} {\rm log} G(x) \right|_* \, ,
\end{equation}
and $\mu$ is the corresponding derivative for $L$. 

The advantage of writing the Jacobian in this way is that all the parameters that appear have an intuitive interpretations. The parameter $\alpha$ is a turnover rate and defines the timescale of the system. The parameters $\mu$ and $\gamma$ are logarithmic derivatives, also called elasticities, which have advantageous statistical properties\cite{Nievergelt1983}. Moreover, for any power law, the corresponding elasticity is the exponent of the power law, e.g.~linear losses imply $\mu=1$ and quadratic losses imply $\mu=2$. 

The generalized model gives us analytical access to a convenient Jacobian that describes a broad class of systems. For instance, the Jacobian for the small example, Eq.~(\ref{smalljacobian}), shows that in every system of that form, a given steady state is stable if the elasticity of loss in the steady state is greater than the elasticity of gain.   
The generalized food web model extends the sione-species case to a set of populations $X_1,\ldots,X_{M}$ described by
\begin{align}
\dot X_a = 
  & G_a(X_a) - M_a(X_a) \notag + \epsilon_a F_a(X_a,T_a(\boldsymbol{X}))\\ & - \sum_b \frac{R_{ab}(A_a,\boldsymbol{X})}{T_b((\boldsymbol{X})}F_b(X_b,T_a(\boldsymbol{X}))=:z_{a}
    \label{foodgm}
\end{align}
where we introduced $z_a$ an abbreviation that is useful below, and $G_a$, $M_a$, $F_a$ are the gain of species $a$ due to primary production, the non-predatory mortality of species $a$, and the total gain of species $a$ from predation, respectively. The function $R_{ab}$ is the amount of species $a$ that is effectively available to predator of species $b$. Often this is a linear function of $X_a$, where the constant of proportionality is describes depends on the ability of species $a$ to capture individuals of species $b$. Finally, 
\begin{equation}
T_a=\sum_b R_{ba}
\end{equation}
is the total amount of prey effectively available to species $a$.

The Eq.~(\ref{foodgm}) can be normalized along the lines laid out in the simple example. As a result we obtain a $M\times M$ Jacobian matrix $\bf P$ with diagonal entries
\begin{align}
\label{eqn:P_diagonal}
	P_{aa}& =  \alpha_a
	\Bigg[
	   \tilde \nu_a \tilde \delta_a \phi_a \notag 
		+ \tilde \nu_a \delta_a \left( \gamma_a \chi_{aa} \lambda_{aa} + \psi_a \right) 
		- \tilde \rho_a \tilde \sigma_a \mu_a \notag \\
		- &\tilde \rho_a \sigma_a
		\Bigg(
			\beta_{aa} \psi_a
			 + \sum_c \beta_{ca} \lambda_{ca}
			\left[
				\left(\gamma_c-1 \right)\chi_{ca}+1
			\right]
		\Bigg)
	\Bigg]
\end{align}
and nondiagonal entries
\begin{align}
\label{eqn:P_non-diagonal}
	P_{ab}& =  
	\alpha_a
	\Bigg[
		 \tilde \nu_a \delta_a \gamma_a \chi_{ab} \lambda_{ab} \notag \\
		& - \tilde \rho_a \sigma_a
		\left(
			\beta_{ba} \psi_b + \sum_c \beta_{ca}\lambda_{cb} \left( \gamma_c - 1 \right) \chi_{cb}
		\right)
	\Bigg],
\end{align}
where the parameters appearing in these equations are elasticities and turnover parameters describing the biomass flow in the system. One can now either estimate these parameters for a given experimental system \cite{Yeakel2014a} or one can use the generalized model to generate plausible random food webs \cite{Plitzko2012}. Here we use the second alternative. 

We use the so-called Niche Model to generate realistic food web topologies \cite{Williams2000}. This model randomly assigns a body mass to every species and then determines the feeding interactions based on these body masses. The generalized model parameters are then drawn from suitable distributions that are dependent on the position of the species in the food web and the body mass parameter. In this way realistic feeding behavior and realistic so-called allometric scaling of certain parameters with the body mass can be incorporated in the model. In the past considerable effort has gone into investigating the realistic ranges and distributions \cite{Plitzko2012} the full parameters sets used in this paper are reproduced in the appendix, and longer discussions of the parameters and their interpretation can be found in \cite{Gross2006,Yeakel2011,Plitzko2012}.

In the present paper we extend the generalized food web model to a meta-foodweb context. We consider a system consisting of $N$ distinct habitat patches, where the dynamics within each patch $i$ is given by the right-hand side of Eq.~(\ref{foodgm}), abbreviated by $z_a$. Additionally the populations are subject to spatial dispersal, modeled as a diffusion process on the network.

For clarity we now use superscript indices $i,j$ to denote the spatial patch while we continue to use subscript indices $a,b,c$ to denote the species. Using this convention we can write the equations of motion as   
\begin{align}
	\dot X_a^i = z_a(\boldsymbol{X^i}) + \sum_j  \left(E_{a}^{ij}(\boldsymbol{X^i}) - E_{a}^{ji}(\boldsymbol{X^i}) \right)  
\label{gma}
\end{align}
where $E_a^{ij}$ is the emigration rate of  individuals of species $a$ from patch $j$ to patch $i$. 
These equations constitute the generalized meta-foodweb model that is our example system in this paper. 

By normalizing, linearizing and then identifying elasticities and turnover rates we can express the Jacobian matrix as a function of interpretable parameters. We note that this procedure can be applied to the reaction and the diffusion part of the equation independently. 
For the reaction part the procedure is completely analogous to non-spatial generalized food web model that we discussed above. For the diffusion part the treatment is analogous to the example of the reaction-diffusion system in Sec.~\ref{secDiffusion}. In particular, the Matrix ${\bf C}$ is obtained by taking the derivative of the (normalized) emigration term with respect to the (normalized) concentrations, 
\begin{equation}
A^{ij}C_{ab}=\left.\frac {\partial{\rm log} E_a^{ij}} {\partial {\rm log}X_b^i}\right|_* 
\label{defC}
\end{equation}
with the adjacency matrix ${\bf A}$. For normal diffusion ${\bf C}$ is a diagonal matrix where $C_{aa}$ is the diffusion constant of species $a$, however for more complex cases such as cross-diffusion, where predators leave preferentially if prey is low or prey flees if there are too many predators, it contains non-diagonal terms (see Sec.~\ref{secDiffusion}).
We present the detailed derivation in the appendix.

In summary, the $NM\times NM$-dimensional Jacobian matrix of the generalized meta-foodweb model can be written in the form Eq.~(\ref{jacobian})
\begin{equation}
\mathbf{J}=\mathbf{I} \otimes \mathbf{P} - \mathbf{L} \otimes \mathbf{C}\, .
\label{jacobianwithc}
\end{equation}
 For the present model the matrix $\bf P$ is the Jacobian matrix of the non-spatial meta-foodweb model (Eqs.~(\ref{eqn:P_diagonal}), (\ref{eqn:P_non-diagonal})),
 $\bf L$ is the (possibly weighted) Laplacian of the underlying geographical network, and $\bf C$  is given by Eq.~(\ref{defC}). (See also Fig.~1 for illustration.) 

\section{Diffusion-driven instabilities in Meta-Foodwebs \label{secResults}}
The previous section showed that the generalized meta-foodweb model falls in the class of systems to which the results from Sec.~\ref{secDDI} apply. We can thus compute the eigenvalues using the formula in Eq.~(\ref{qlambda}).

\begin{figure*}[t]
\includegraphics[width=\textwidth]{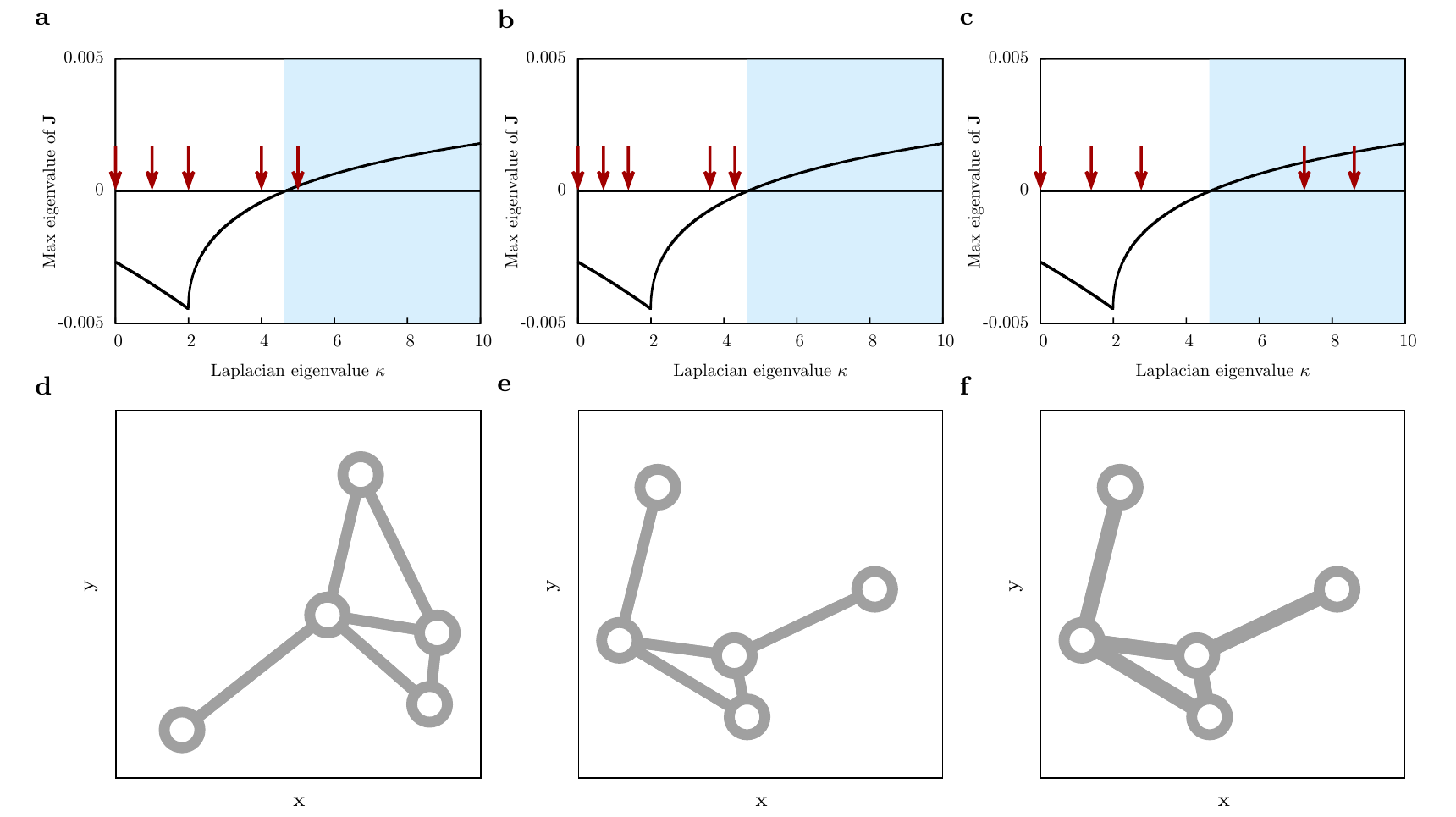}  
\caption{Master Stability Function (MSF) for meta-foodwebs. Shown are MSFs (a-c) and spatial geometries (d-f) for the same local food web (that of Fig.~1) and coupling matrix $\mathbf{C}$ (see Eq.~(\protect\ref{jacobianwithc}) and Supplementary Material for details) but 3 different geometries (represented by circles, indicating patches, and lines, indicating migration links, in a two dimensional x,y landscape).  
The MSF (identical line in a-c) relates the Jacobian eigenvalue of the meta-foodweb to the Laplacian eigenvalue of the spacial network. The meta-foodweb is stable if none of the Laplacian eigenvalues (arrows) fall into ranges where the master stability function (line in a-c) is positive (blue shaded area). The food web under consideration is unstable on one geometry (a,d) but stable on another (b,e). Tightening the coupling, i.e.~increasing the weights in the adjacency matrix  (indicated by thicker lines in f),  can destabilize the stable system by stretching the Laplacian spectrum (cf. b,c).} 
\end{figure*}

Equation \eqref{qlambda} has the useful property that the structure of the spatial networks only enters through the eigenvalues. One can say that every Laplacian eigenvalue $\kappa_i$ generates a set of Jacobian eigenvalues that is independent of the other Laplacian eigenvalues.
This means that Eq.~\eqref{qlambda} defines a master stability function:  Given only information about the local system (i.e.~${\bf P}$ and ${\bf C}$), we can compute the leading eigenvalue $\lambda_{\rm max}$ that would be generated by a given Laplacian eigenvalue $\kappa$ \cite{Pecora1998,Arenas2008}. 
The resulting function $S(\kappa)={\rm Re}(\lambda_{\rm max}(\kappa))$ is then a master stability function for the meta-foodweb under consideration. 

Because stability requires all eigenvalues of the Jacobian to have negative real parts, stability is lost if any Laplacian eigenvalue falls into a range where the master stability function is positive. In the following we refer to these ranges as ``forbidden'' as they have to be avoided if local stability is to be maintained. The loss of stability that occurs when eigenvalues enter these regions is analogous to the onset of pattern formation (Turing bifurcations, and wave instabilities) in continues space. 

\subsection{Stability of small food webs}
Let us first consider the stability properties of the  the 4-species foodweb from Fig.~1. The master stability function corresponding to this food web is plotted in Fig.~3. Because the function only depends on the local network and the nature of the coupling, it is independent of the underlying spatial network into which the foodweb is placed. However, different spatial networks have different Laplacian spectra and thus sample the master stability function at different points, leading to different stability properties.  

Linking the stability properties of the food web to Laplacian eigenvalues is interesting because the dependence of Laplacian spectrum on the topology of the network is relatively well understood \cite{Merris1998}. The master stability function thus offers an opportunity to understand how the stability of network reaction-diffusion systems, or its loss, depends on topological properties.

The Laplacian matrix is a positive semidefinite matrix that always contains at least one zero eigenvalue. Hence the spectrum of $\bf J$ always contains the set of eigenvalues generated by $\kappa=0$, which are also the eigenvalues of $\bf P$. This shows that the homogeneous state in the meta-foodweb can only be stable if the corresponding steady state in the non-spatial food web is stable. 

We numerically computed the MSF for a variety of randomly generated food webs. 
In smaller webs with up to 5 species we mostly observed MSFs with relatively simple shapes, where the MSF is either (i) positive at zero, (ii) negative everywhere, or (iii) crosses from negative to positive values at a single $\kappa^*>0$ (cf.~Fig.~3). 
The former two cases correspond to food webs that are unstable (i) or stable (ii) irrespective of the geographical network, whereas the third case (iii) is stable if all eigenvalues of the Laplacian are sufficiently small ($\kappa_i<\kappa^*$).
 
Uniformly increasing the diffusion constant in a given network stretches the Laplacian spectrum. In systems of class (iii) the homogeneous state is therefore at greater risk of instability in networks where the diffusive coupling is stronger. This shows that densely linked landscapes, which are thought to be beneficial in ecology, may lead to instability of the homogeneous steady state in a food webs of class (iii). However, this instability is not necessarily detrimental as it can lead to stable spatial patterns that introduce heterogeneity, which might ultimately benefit the diversity of the system\cite{Stein2014}.    

\begin{figure*}[t]
\includegraphics[width=0.95\textwidth]{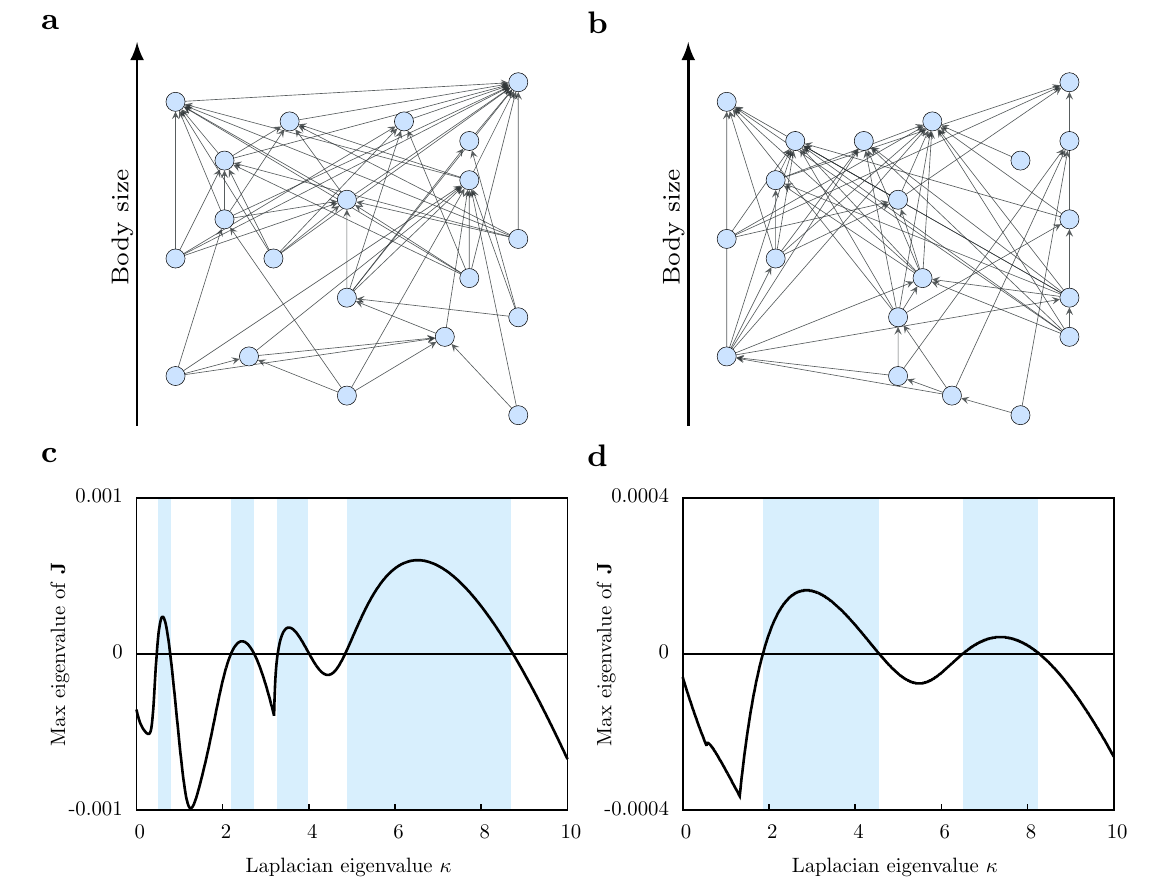}  
\caption{Complex Master Stability Functions (MSFs). Shown are two examples of foodwebs (a,b) of 20 species (blue bubbles) connected by predator-prey interactions (arrows). The coupling matrix $\mathbf{C}$ was constructed such that predators emigrate preferentially from patches with scarce prey and prey emigrates preferentially from patches with abundant predators (see Supplementary Material). The corresponding MSFs (c for a, d for b) have many forbidden (blue) ranges.}
\end{figure*}

\begin{figure*}[t]
\includegraphics[width=\textwidth]{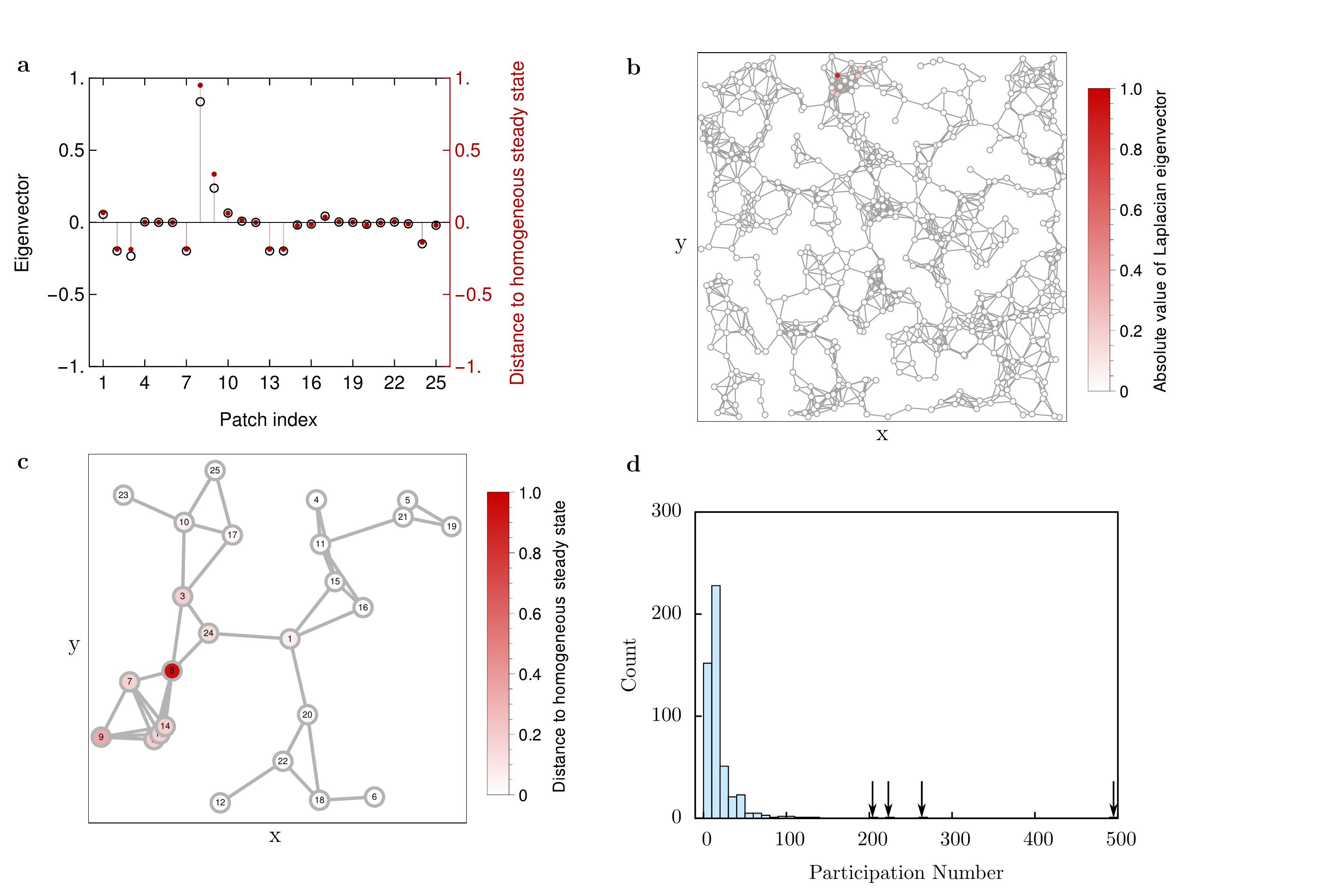} 
\caption{Localization of Eigenvectors. Shown is a system where on a given topology (c) one eigenvalue of the Laplacian 
lies in one of the forbidden ranges where the MSF of the food web is positive. Hence the system departs from the homogeneous steady state and approaches a non-homogeneous state. Color-coding the distance of the density of the apex predator population to the homogeneous value (see methods) shows that the impact of the instability is confined to a relatively small area. In cases where the system approaches a state in the vicinity of the homogeneous state, the eigenvector corresponding to the destabilizing eigenvector is indicative of the pattern in the final state. A comparison (a) of the respective eigenvector (open circles) and the observed deviation from the homogeneous value in the final state (dots) shows good agreement in the example system (c), but similar accuracy cannot be guaranteed in general. The localization of many eigenvectors is a generic feature of geographical networks. For the 500-node example geometry in b the number of nodes on which eigenvectors have a significant amplitude can be quantified by the participation number (see methods). This reveals that the majority of eigenvectors only extend to relatively few nodes (histogram in d), while only 4 eigenvalues (arrows) have significant amplitude on a large fraction of the nodes. The example shown here is a case of real eigenvalues causing a Turing instability. When the leading eigenvalue is complex, the associated instability is a wave instability (see Supplementary Material).}
\end{figure*}

\subsection{Stability of larger foodwebs}
The analysis of small food-webs, including the example in Fig.~3 revealed relatively simple master stability functions, which are very similar to the master stability functions observed in coupled oscillator models. However, in large food webs and particularly if cross-diffusion is allowed, much more complex functions can be observed. 

Figure 4 shows two examples of 30-species meta-foodwebs and their corresponding master stability functions. The functions go through a complex pattern of stable and unstable intervals that defies easy classification. One could now explore the transitions in which forbidden regions are created and vanish, but this analysis exceeds the scope of the present paper.

For ecology the complex master stability functions are sobering as they show that there cannot be any simple laws that govern the stability of these systems. The only exception is perhaps that we can say that in a sufficiently weakly coupled system the Laplacian eigenvalues always cluster around zero. Since the master stability function has to be a continuous function there will always be a critical threshold for the coupling strength below which the stability properties of the weakly coupled system are identical to the stability properties of an isolated system. 

In absence of easy rules for the stability of meta-foodwebs  ecologists will have to consider different food webs and different spatial networks individually. We believe that the methodology proposed here will prove conducive to this task.  

\subsection{Network spectra and localized modes}
Throughout most of this paper we have argued that extensive analogies exist between reaction-diffusion systems in space and reaction-diffusion systems on networks. However differences exist in the spectrum of the respective Laplacian operators. 

In continuous spatial domains the spectrum of the Laplacian operator tends to be relatively regular and the eigenmodes are typically delocalized functions such as trigonometric or Bessel functions. By contrast the spectrum of the Laplace operator on networks has a complex structure, and the majority of eigenmodes are localized \cite{Cucuringu2011,Nyberg2014,Dettmann2017}. 

For illustration we consider the well-studied example of random geometric graphs \cite{Dall2002}, which provide a generic model for spatial networks. They consists of nodes that are randomly placed in a square and that are connected to all neighbors within a certain distance. The extent to which an eigenvector is localized can be quantified by the participation number \cite{Nyberg2014} which indicates the approximate number of nodes on which the eigenvector has a significant amplitude.  
In a example system of 500 spatial patches we find that more than 150 modes have a participation number of less than 10 and more than 200 additional modes have a participation number between 10 or 20 (Fig.~5). Thus more than half of the eigenvectors are localized on small clusters that contain less than 4\% of the nodes.  

In the context of reaction-diffusion systems localized spatial modes have two implications
\begin{itemize}
\item The corresponding eigenvalue is only sensitive to the structure inside the cluster, and thus the eigenvalue is informative of local features in that cluster.  
\item If the eigenvalue lies in one of the forbidden regions, such that it generates a dynamical instability, this instability will at least initially be confined to the cluster on which the corresponding eigenmode is localized. 
\end{itemize}

These properties of networks may make it possible to engineer systems where the dynamics reacts sensitively to local topological properties,  a possibility that we discuss briefly in the conclusions. 

\subsection{Numerical validation and behavior in unstable regions}
We checked the theoretical results numerically by numerical diagonalization of the Jacobian matrix and by simulating specific dynamical models, which yielded 
a perfect agreement. Furthermore, we noticed that in systems where the homogeneous state is unstable the system often approaches heterogeneous steady state for which variables in the different nodes are closely predicted by the unstable eigenmode. 

While the mathematics imply that the system leaves the unstable state in a direction corresponding to the leading eigenvector\cite{Guckenheimer1983}, we are not aware of a mathematical reason why the eigenvector should remain informative after this initial departure. In low-dimensional systems one typically observes that the system approaches some other attractor that is far away from the unstable state. While such catastrophic departures also occur in the network reaction-diffusion systems studied here, we also observed a large proportion of simulation runs where the developing pattern closely resembled the unstable eigenmode 
(see Fig.~5 for an example, and video linked in the appendix).    

\section{Conclusions \label{secConclusions}}
In this paper we have shown that a deep analogy exists between reaction-diffusion system in continuous space and reaction-diffusion system on networks. The analysis of pattern-forming diffusion-driven instabilities in networks is thus not more difficult that the analysis of such instabilities in continuous space; perhaps even easier as eigenmodes of large networks are easier to compute than eigenfunctions of complex spatial domains.
Yet networks have often a more complex spectrum including localized modes which can give rise to localized patterns. 

Our analysis, while relying on Kronecker products, leads to a formulation that constitutes a master stability function. Given the long tradition that the master stability approach has in synchronization and the many modern applications, it is surprising that the application to stationary states proposed here has not been used in a wide range of applications already. In our opinion there is plenty of untapped potential to leverage this approach for progress in applications.

Reaction-diffusion systems on networks can be described as multi-layer networks, a class of systems that has recently gained much attention. Compared to other multi-layer systems the networks studied here fall into a comparatively tame class as the layers are topologically identical. However, even with this restriction network reaction-diffusion systems can accommodate a high degree of complexity. We illustrated this point by analyzing the generalized meta-foodweb model. This model described a set of species which have different nonlinear dynamics reflecting their different biology. 

Although we required that all populations disperse over the same spatial network, diffusion can take place at different rates, which can moreover depend dynamically on the local conditions. Our analysis of small foodwebs revealed that in a broad class of system, tight spatial coupling can destabilize the homogeneous stationary states leading to pattern formation. 

We also showed that large food webs can have very complex master stability functions, such that there cannot be any easy laws governing the stability of homogeneous states in these systems. This implies that systems from applications have to be studied separately in detail, and thus efficient approaches, such as the one proposed here, are needed.

We emphasize that the approach proposed here is not limited to the analysis of food webs. Let us therefore conclude by outlining other possible applications in biology:
\begin{itemize}
\item In game theory, network effects have been prominently discussed for some time. Using the proposed approach population dynamical models, say of cooperator and defector populations, could be studied to determine under which conditions the state where agents are distributed homogeneously across the network becomes unstable creating patterns where havens for cooperators exist. 
\item In epidemiology stability analysis of homogeneous disease free states have been proposed as a method to compute epidemic thresholds for a single pathogen. However, there is a significant interest in co-infection models using multiple microbes. Using the proposed approach, master stability functions for the invasion of large microbial communities could be computed, providing insights into transmission of microbiomes. In this application the adjacency matrix replaces the Laplacian as the spatial operator, but the approach can be straightforwardly extended to accommodate this matrix. 
\item In cell biology the approach could be used to investigate the dynamics tissues of identical cells. Particularly one could ask how metabolic or signaling interactions could lead to pattern formation. In the context synthetic biology the abundance of localized eigenmodes in spatial networks might be exploited to engineer reaction systems for which pattern formation localizes on certain topological features, e.g.~detecting for example areas in the tissue where cells are particularly highly clustered.   
\end{itemize}

We hope that the approach proposed here will be conducive to advancing the understanding in these and other applications. \\

\small{
{\bf Acknowledgements. }
We thank Dr. Lars Rudolf for help with the generation of generalized niche model webs. This work was supported by DFG projects number Dr300/12-2 and Dr300/13-2 and EPSRC projects EP/K031686/1, and EP/N034384/1. It was also supported in part by Perimeter Institute for Theoretical Physics. Research at Perimeter Institute is supported by the Government of Canada through the Department of Innovation, Science and Economic Development Canada and by the Province of Ontario through the Ministry of Research, Innovation and Science.

{\bf Data Statement. }
This work did not produce or use primary research data. Details of the models can be found in the supplementary material.
}


\bibliography{metafoodweb}
\bibliographystyle{apsrev-title}

\pagebreak ~\pagebreak

\section*{Appendix}

\subsection{Full generalized meta-foodweb model}
We denote the biomass density of species $i$ in habitat $k$ by $X_i^k$. Its change is given by
\begin{align}
\label{eqn:biomass_ode}
\dot X_i^k
	&= G_i^k(X_i^k) - M_i^k(X_i^k) \notag \\
	&+ \epsilon_i F_i^k(X_1^k, \ldots ,X_S^k) - \sum_j D_{ji}^k(X_1^k,\ldots,X_S^k) \notag \\
	&+ \sum_l \Big[ E_i^{kl}(X^k_1,\ldots,X^k_S,X^l_1,\ldots,X_S^l) \notag \\
	&\qquad\qquad - E_i^{lk}(X^l_1,\ldots,X^l_S,X^k_1,\ldots,X_S^k) \Big] \;,
\end{align}
where $G_i^k$ is the growth by primary production, $M_i^k$ is the loss by respiration and mortality, $F_i^k$ is the growth due to predation, $D_{ji}^k$ is the loss of biomass due to predation by species $j$, $E_i^{kl}$ is the migration from habitat $l$ to $k$. Furthermore we used $\epsilon_i$ to denote the conversion efficiency of prey biomass.

Following Gross~\textit{et al.}~\cite{Gross2006} we capture the correlation between the loss of the prey species $D_{ji}^k$ and the growth of the corresponding predators $F_j^k$, by introducing the auxiliary variable for the  total amount of prey that is available to species $j$ in habitat $k$,
\begin{align}
	T_j^k (X_1^k,\ldots,X_S^k) = \sum_i R_{ji}^k(X_i^k) \;,
\end{align}
where $R_{ji}^k(X_i^k)$ is the relative contribution of species $i$ in patch $k$ to the total amount of food available to the population of species $j$ in the habitat. 

We can now write the amount of prey consumed by population $j$ as
\begin{align}
	F_j^k(X_1^k,\ldots,X_N^k) = F_j^k(T_j^k,X_j^k) \;,
\end{align}
and the loss of species $i$ due to predation by species $j$ in habitat $k$ as
\begin{align}
	D_{ji}^k(X_1^k,\ldots,X_S^k) = \frac{R_{ji}^k(X_i^k)}{T_j^k (X_1^k,\ldots,X_S^k)} F_j^k(T_j^k,X_j^k) \;.
    \label{eqn:relative_loss}
\end{align}

\subsection{Derivation of Jacobian Matrix}
We assume that the system in eqn.~\eqref{eqn:biomass_ode} has at least one positive, (but potentially unstable) steady state. This is a very mild assumption, given the considerable freedom that still exists in the class of models, 
We then denote the (unknown) steady-state population densities by $X_i^{k*}$. Likewise we use the asterisk to denote functions evaluated in the steady state, e.g $F_i^{k\ast}=F_i^{k}(X_1^{k*},\ldots,X_S^{k*} )$. 
We then normalize all dynamical variables and functions by their steady-state value. The results of this normalization are denoted by lowercase symbols  For instance
$ x_i^k = \frac{X_i^k}{X_i^{k*}} $.

Using these definitions we obtain the normalized equations
\begin{align}
	\dot x_i^k &=
	\frac{G_i^{k*}}{X_i^{k*}} g_i^k(x_i^k) - \frac{M_i^{k*}}{X_i^{k*}} m_i^k(x_i^k) \notag \\
	&+ \frac{\epsilon_i F_i^{k*}}{X_i^{k*}} f_i^k(t_i^k,x_i^k) \notag \\
	&- \sum_j \frac{D_{ji}^{k*}}{X_i^{k*}} d_{ji}^k(x_1^k,\ldots,x_S^k) \notag \\
	&+ \sum_l \Bigg[ \frac{E_i^{kl*}}{X_i^{k*}} e_i^{kl}(x^k_1,\ldots,x^k_S,x^l_1,\ldots,x_S^l) \notag \\
	&\qquad\qquad - \frac{E_i^{lk*}}{X_i^{k*}} e_i^{lk}(x^l_1,\ldots,x^l_S,x^k_1,\ldots,x_S^k) \Bigg] \;.
\end{align}

We can now identify a set of structural parameters that characterize the biomass flow in the steady state under consideration. In the context of GM such parameters are called \emph{scale parameters}.  We start with the time scales
\begin{align}
	\alpha_i^k &=
	\frac{G_i^{k*}}{X_i^{k*}} + \frac{\epsilon_i F_i^{k*}}{X_i^{k*}} + \sum_l \frac{ E_i^{kl*}}{X_i^{k*}} \notag \\
	&= \frac{M_i^{k*}}{X_i^{k*}} + \sum_j \frac{D_{ji}^{k*}}{X_i^{k*}} + \sum_l \frac{E_i^{lk*}}{X_i^{k*}} \;.
\end{align}
These scale parameters quantify the rate of biomass flow in the steady state. The relative contributions to the biomass gain by the different processes are
\begin{align}
	\nu_i^k &= \sum_l \nu_i^{kl} = \frac{1}{\alpha_i^k} \sum_l \frac{ E_i^{kl*}}{X_i^*} \;, \\
	\tilde \nu_i^k &= 1 - \nu_i^k = \frac{1}{\alpha_i^k} \frac{\epsilon_i F_i^{k*}}{X_i^{k*}} + \frac{1}{\alpha_i^k} \frac{G_i^{k*}}{X_i^{k*}} \;, \\
	\tilde \nu_i^k \delta_i^k &= \frac{1}{\alpha_i^k} \frac{\epsilon_i F_i^{k*}}{X_i^{k*}} \;, \\
	\tilde \nu_i^k \tilde \delta_i^k &= \tilde \nu_i^k (1-\delta_i^k) = \frac{1}{\alpha_i^k} \frac{G_i^{k*}}{X_i^{k*}} \;.
\end{align}
The relative contributions of the different processes to the biomass loss are given by
\begin{align}
	\rho_i^k &= \sum_l \rho_i^{lk} = \frac{1}{\alpha_i^k} \sum_l \frac{E_i^{lk*}}{X_i^*} \;, \\
	\tilde \rho_i^k &= 1 - \rho_i^k = \frac{1}{\alpha_i^k} \frac{M_i^{k*}}{X_i^{k*}} + \frac{1}{\alpha_i^k} \sum_j \frac{D_{ji}^{k*}}{X_i^{k*}} \;, \\
	\tilde \rho_i^k \sigma_i^k &= \frac{1}{\alpha_i^k} \sum_j \frac{D_{ji}^{k*}}{X_i^{k*}} \;, \\
	\tilde \rho_i^k \tilde \sigma_i^k &= \tilde \rho_i^k (1 - \sigma_i^k) = \frac{1}{\alpha_i^k} \frac{M_i^{k*}}{X_i^{k*}} \;.
\end{align}
It is necessary to resolve the contribution of different species to the loss by additionally defining the parameters
\begin{align}
	\beta_{ji}^k &= \frac{1}{\alpha_i^k \tilde \rho_i^k \sigma_i^k} \frac{D_{ji}^{k*}}{X_i^{k*}} \;.
\end{align}
Using eqn. \eqref{eqn:relative_loss} the normalized function for the loss due to predation can be written as
\begin{align}
	d_{ji}(x_1^k, \ldots ,x_S^{k}) &= \frac{R_{ji}^{k*} F_j^{k*}}{T_j^{k*} D_{ji}^{k*}} \frac{r_{ji}^k}{t_j^k} f_j(t_j^k, x_j^k) \notag \\
	&= \frac{r_{ji}^k}{t_j^k} f_j(t_j^k, x_j^k) \;,
\end{align}
where the normalized total available biomass for predation by species $j$ is given by
\begin{align}
	t_j^k = \sum_i \frac{R_{ji}^{k*}}{T_j^{k*}}r_{ji}^k \;.
\end{align}
Using the parameters
\begin{align}
	\chi_{ji}^k = \frac{R_{ji}^{k*}}{T_j^{k*}},
\end{align}
we can write
\begin{align}
	t_j^k &= \sum_i \chi_{ji}^k r_{ji} \;.
\end{align}
In summary this yields the normalized meta-foodweb model
\begin{align}
	\dot x_i^k = \alpha_i^k
	\Bigg[
		&\quad \; \tilde \nu_i^k \tilde \delta_i^k g_i^k(x_i^k) \notag\\
		&+ \tilde \nu_i^k \delta_i^k f_i^k(t_i^k,x_i^k) \notag\\
		&- \tilde \rho_i^k \tilde \sigma_i^k m_i^k(x_i^k) \notag\\
		&- \tilde \rho_i^k \sigma_i^k \sum_j \beta_{ji}^k d_{ji}^k(x_1^k,\ldots,x_S^k) \notag\\
		&+ \sum_l \nu_i^{kl} e_i^{kl}(x^k_1,\ldots,x^k_S,x^l_1,\ldots,x_S^l) \notag\\
		&- \sum_l \rho_i^{lk} e_i^{lk}(x^l_1,\ldots,x^l_S,x^k_1,\ldots,x_S^k) 
	\Bigg] \;, \label{eqn:normalized_model}
\end{align}
where $i=1,\ldots,S$ and $k=1,\ldots,N$.

\subsection{Calculation of the Jacobian}
Our model still contains unknown functional forms. However the only aspect of this uncertainty that is relevant for the local dynamics are certain derivatives of the functions evaluated in the unknown steady state under consideration. The core of GM is the idea that this uncertainty can be captured in so called \emph{exponent parameters}, which we define as 
\begin{align}
	\phi_i^k &= \left. \frac{\partial}{\partial x_i^k} g_i^k(x_i^k) \right|_{x=x^*} \;, \\
	\mu_i^k &= \left. \frac{\partial}{\partial x_i^k} m_i^k(x_i^k) \right|_{x=x^*} \;, \\
	\lambda_{ji}^k &= \left. \frac{\partial}{\partial x_i^k} r_{ji}^k(x_i^k) \right|_{x=x^*} \;, \\
	\gamma_i^k &= \left. \frac{\partial}{\partial t_i^k} f_i^k(t_i^k,x_i^k) \right|_{x=x^*} \;, \\
	\psi_i^k &= \left. \frac{\partial}{\partial x_i^k} f_i^k(t_i^k,x_i^k) \right|_{x=x^*} \;,
\end{align}
and for migration
\begin{align}
	\hat \omega_i^{kl} &= \left. \frac{\partial}{\partial x_i^k} e_i^{kl}(x^k_1,\ldots,x^k_S,x^l_1,\ldots,x_S^l) \right|_{x=x^*} \;, \\
	\omega_i^{kl} &= \left. \frac{\partial}{\partial x_i^l} e_i^{kl}(x^k_1,\ldots,x^k_S,x^l_1,\ldots,x_S^l) \right|_{x=x^*} \;, \\
	\hat \kappa_{ij}^{kl} &= \left. \frac{\partial}{\partial x_j^k} e_i^{kl}(x^k_1,\ldots,x^k_S,x^l_1,\ldots,x_S^l) \right|_{x=x^*} \; \text{mit } i \neq j \;, \\
	\kappa_{ij}^{kl} &= \left. \frac{\partial}{\partial x_j^l} e_i^{kl}(x^k_1,\ldots,x^k_S,x^l_1,\ldots,x_S^l) \right|_{x=x^*} \; \text{mit } i \neq j \;.
\end{align}

\subsection{Matrix Representation}
We use three types of matrices to construct the Jacobian. The matrices $\mathbf P^k$ capture the local biology within a patch, whereas the matrices $\mathbf C^{kl}$ and $\mathbf{\hat C}^{kl}$ capture the dependencies caused by migration from patch $l$ to $k$. From the GM above we find the diagonal elements of $\mathbf P^k$ 
\begin{align}
	P_{ii}^k = \notag \\
	\alpha_i^k
	\Bigg[
		&\quad \; \tilde \nu_i^k \tilde \delta_i^k \phi_i^k \notag \\
		&+ \tilde \nu_i^k \delta_i^k \left( \gamma_i^k \chi_{ii}^k \lambda_{ii}^k + \psi_i^k \right) \notag \\
		&- \tilde \rho_i^k \tilde \sigma_i^k \mu_i^k \notag \\
		&- \tilde \rho_i^k \sigma_i^k
		\Bigg(
			\beta_{ii}^k \psi_i^k
			 + \sum_n \beta_{ni} \lambda_{ni}
			\left[
				\left(\gamma_n^k-1 \right)\chi_{ni}^k+1
			\right]
		\Bigg)
	\Bigg] \;,
\end{align}
and the non-diagonal elements
\begin{align}
	P_{ij}^k = \notag \\
	\alpha_i^k
	\Bigg[
		&\quad  \tilde \nu_i^k \delta_i^k \gamma_i^k \chi_{ij}^k \lambda_{ij}^k \notag \\
		&- \tilde \rho_i^k \sigma_i^k
		\left(
			\beta_{ji}^k \psi_j^k + \sum_n \beta_{ni}^k\lambda_{nj}^k \left( \gamma_n^k - 1 \right) \chi_{nj}
		\right)
	\Bigg] \;.
\end{align}
The matrices $\mathbf C^{kl}$ have diagonal elements
\begin{align}
	C_{ii}^{kl} &= \alpha_i^k
	\left[
        \rho_i^{lk} \omega_i^{lk}
		- \nu_i^{kl} \hat \omega_i^{kl}
	\right]
\end{align}
and non-diagonal elements
\begin{align}
	C_{ij}^{kl} &= \alpha_i^k
	\left[
     	\rho_i^{lk} \kappa_{ij}^{lk}
		- \nu_i^{kl} \hat \kappa_{ij}^{lk}
	\right] \;.
\end{align}
The analogously defined matrices $\mathbf{\hat C}^{kl}$ with swapped scale parameters have the diagonal elements
\begin{align}
	{\hat C}_{ii}^{kl} &= \alpha_i^k
	\left[
    	\nu_i^{lk} \omega_i^{lk}
		- \rho_i^{kl} \hat \omega_i^{kl}
	\right]
\end{align}
and the non-diagonal elements
\begin{align}
	{\hat C}_{ij}^{kl} &= \alpha_i^k
	\left[
    	\nu_i^{lk} \kappa_{ij}^{lk}
		- \rho_i^{kl} \hat \kappa_{ij}^{kl}
	\right] \;.
\end{align}
 The Jacobian is then constructed as
\begin{align}
	\mathbf J = \begin{pmatrix}
		\ddots \\
		& \mathbf P^k - \sum_m \mathbf C^{km} & \cdots &          \mathbf {\hat C}^{kl} \\
		& \vdots & \ddots & \vdots \\
		& \mathbf {\hat C}^{lk} & \cdots & \mathbf P^l - \sum_m \mathbf C^{lm} \\
		&&&& \ddots
	\end{pmatrix} \;.
\end{align}
The unknown stable steady state under consideration is stable if all eigenvalues have negative real parts. 

Although the functions and the steady states are unknown, the exponent and scale parameters that enter the Jacobian are directly interpretable in the context of the system and can therefore be measured directly in nature or can be chosen to based on theoretical considerations (see Gross \textit{et al.}~\cite{Gross2006,Gross2009} and table below for details). 

Note that different steady states are characterized by different values of the scale and exponent parameters and thus can have different stability properties. 

In the context of this paper we consider the stability of homogeneous steady states, i.e.~states in which all patches are characterized by the same parameters and have the same biomass density. We then speak off diffusion-driven instability (DDI) if such a state is stable without diffusion, but loses stability under  non-zero diffusive coupling.   

\subsection{Diffusive Mass Balance in Homogeneous States}
For a homogeneous equilibrium the incoming flux experienced by a given population must equal the  outgoing flux
\begin{align}
	\tilde \nu_i^k \alpha_i^k  &= \tilde \rho_i^k \alpha_i^k \;
\end{align}
and furthermore
\begin{align}
	\nu_i^k \alpha_i^k  &= \rho_i^k \alpha_i^k \;.
\end{align}
Therefore we obtain
\begin{align}
	\tilde \nu_i^k &= \tilde \rho_i^k
\end{align}
and
\begin{align}
	\nu_i^k &= \rho_i^k \;.
\end{align}
Thus we define the local biomass flow
\begin{align}
	{\alpha_P}_i^k = \tilde \nu_i^k \alpha_i^k = \tilde \rho_i^k \alpha_i^k \;
\end{align}
and the biomass flow due to migration
\begin{align}
	{\alpha_C}_i^k = \nu_i^k \alpha_i^k = \rho_i^k \alpha_i^k \;.
\end{align}
At this point we have to introduce the auxiliary scale parameters $\eta_i^{kl}$ to keep track of the relative contributions of the spatial links to the migration bimass flow:
\begin{align}
	\nu_i^{kl} \alpha_i^k = \rho_i^{kl} \alpha_i^k = \eta_i^{kl} {\alpha_C}_i^k \;.
\end{align}

Using these definitions we can rewrite the matrices $\mathbf P^k$ and $\mathbf C^{kl} = \mathbf{\hat C}^{kl}$. The matrix for local dynamics is given by
\begin{align}
    \label{eqn:P_diagonal2}
	P_{ii}^k = \notag \\
	{\alpha_P}_i^k
	\Bigg[
		&\quad \;  \tilde \delta_i^k \phi_i^k \notag \\
		&+ \delta_i^k \left( \gamma_i^k \chi_{ii}^k \lambda_{ii}^k + \psi_i^k \right) \notag \\
		&- \tilde \sigma_i^k \mu_i^k \notag \\
		&- \sigma_i^k
		\Bigg(
			\beta_{ii}^k \psi_i^k
			 + \sum_n \beta_{ni} \lambda_{ni}
			\left[
				\left(\gamma_n^k-1 \right)\chi_{ni}^k+1
			\right]
		\Bigg)
	\Bigg]
\end{align}
and
\begin{align}
    \label{eqn:P_nondiagonal2}
	P_{ij}^k = \notag \\
	{\alpha_P}_i^k
	\Bigg[
		&\quad \delta_i^k \gamma_i^k \chi_{ij}^k \lambda_{ij}^k \notag \\
		&- \sigma_i^k
		\left(
			\beta_{ji}^k \psi_j^k + \sum_n \beta_{ni}^k\lambda_{nj}^k \left( \gamma_n^k - 1 \right) \chi_{nj}
		\right)
	\Bigg] \;.
\end{align}
The matrix for migration dynamics is given by
\begin{align}
	C_{ii}^{kl} &= {\alpha_C}_i^k
	\left[
		\eta_i^{lk} \omega_i^{lk}
        - \eta_i^{kl} \hat\omega_i^{kl}
	\right] \;,
\end{align}
and
\begin{align}
	C_{ij}^{kl} &= {\alpha_C}_i^k
	\left[
    	\eta_i^{lk} \kappa_{ij}^{lk}
		- \eta_i^{kl} \hat\kappa_{ij}^{kl}
	\right] \;.
\end{align}

\FloatBarrier
\begin{table}
\begin{tabular}{l|p{0.33\textwidth}}
Parameter & Interpretation\\
\hline
Elasticity& \\
\hline
$\phi_i^k$ & Sensitivity of primary production of $i$ in $k$ to itself\\
$\gamma_i^k$ & Sensitivity of predation of $i$ in $k$ to prey density \\   	
$\lambda_i^k$ & Exponent of prey switching of $i$ in $k$\\
$\psi_i^k$ & Sensitivity of predation of $i$ to density of itself\\
$\mu_i^k$ & Exponent of closure of $i$ in $k$\\
$\omega_i^{kl}$ & Sensitivity of migration of $i$ from $l$ to $k$ to itself\\
$\tilde{\omega}_i^{kl}$ & Sensitivity of migration of $i$ from $l$ to $k$ to itself\\
$\kappa_{ij}^{kl}$ & Sensitivity of migration of $i$ from $l$ to $k$ to $j$\\
$\tilde{\kappa}_{ij}^{kl}$ & Sensitivity of migration of $i$ from $l$ to $k$ to $j$\\
\hline
Turnover & \\
\hline
${\alpha_P}_i^k$ & Intra-habitat biomass flow of $i$ in $k$\\
${\alpha_C}_i^k$ &  Biomass flow by migration of $i$ in $k$\\
$\beta_{ji}^k$ & Contribution of predation by $i$ to local biomass loss of $j$\\
$\sigma_i^k$ & Fraction of local biomass loss of $i$ in $k$ due to predation\\
$\tilde{\sigma}_i^k$ & Fraction of local biomass loss of $i$ in $k$ due to respiration\\
$\delta_i^k$ & Fraction of local growth by predation of $i$ in $k$\\  
$\tilde{\delta}_i^k$ & Fraction of local growth by primary production of $i$ in $k$\\  
$\chi_i$ & Contribution of $i$ to prey of $j$ on $k$\\
\hline
$\nu_i^k$ & Fraction of total biomass gain of $i$ in $k$ due to migration\\
$\tilde{\nu}_i^k$ & Fraction of total biomass gain of $i$ in $k$ due to predation\\
$\rho_i^k$ & Fraction of total biomass loss of $i$ in $k$ due to migration\\
$\tilde{\rho}_i^k$ & Fraction of total biomass loss of $i$ in $k$ due to predation\\
$\eta_i^{kl}$ & Fraction of migration biomass flow of $i$ due to migration from $l$ to $k$
\end{tabular}
\caption{Generalized parameters used to describe the meta-foodweb. The indices $i$ and $j$ denote different species and $k$ and $l$ different habitates}
\label{tab:GeneralizedParameters}
\end{table}

\subsection{Niche Model Topologies}
The foodwebs in Figs.~2 and 3 were generated with the niche model \cite{Williams2000}. In this model, each species $i$ is assigned at random a niche value $n_{i}\in[0,1]$, which is related to the biomass turnover rate $\alpha_i$ occurring in the generalized modeling approach.

Each species is assigned a feeding range $r_i$ and a feeding center $c_i$. The feeding range is drawn from a $\beta$ distribution,
\begin{equation}
r_i = \left[1 - (1-x)^{\frac{2C}{1-2C}}\right] \cdot n_i
\end{equation}
with a random number $x\in[0,1]$, and with $C$ being the connectivity of the food-web (See corresponding figure sections for values).

The feeding center $c_i$ is chosen at random from the interval$[\frac r 2, n-\frac{r}{2}]$, and a species $j\neq i$ is a prey of species $i$ if its niche value $n_{j}$ is within the interval $[c_i-r_i/2,c_i+r_i/2]$. 
Every species without a prey species is considered a primary producer. 

Each prey species is assigned a relative contribution to the diet of its predator, which is drawn from a normal distribution. We subsequently normalize the relative contributions such that they add up to 1.  

\subsection{Parameterization of the model}
In the following, we assume that all patches are identical and thus $P_{ij}^k=P_{ij}^l$ for all $k$ and $l$ in $N$. In addition, for simplicity, we assume that the same generalized parameters, which are not based on diet composition or bodymass, have identical values for all species, e.g.  $\phi_i^k = \phi_j^k \equiv \phi$.
Generalized parameters based on diet composition, e.g. $\beta_i^k$, are determined by the generated foodweb. In addition, we assume that biomass flows scale with niche value $n_i$, which is a proxy for bodysize \cite{Kartascheff2010}, i.e $\alpha_{P_i}^k=10^{-2n_i}$ and $\alpha_{C_i}^k=10^{-4n_i}$.
The remaining free parameter for the local foodwebs are, $\phi$, $\gamma$, $\lambda$, $\psi$ and $\mu$. The migration parameters contained in $\mathbf{C}$ are
$\omega$, $\tilde{\omega}$, $\kappa$ and $\tilde{\kappa}$.

\subsection{Details for Figure 3}
For Fig.~3, we generated niche webs with 4 species and a connectivity, $C$, of $0.33\pm 0.01$ until we obtained one with the desired structure shown in Fig.1, with the niche values shown in Table \ref{fig2:niche}.  
Matrix $\mathbf{P}$ is given by equations eqn.~\eqref{eqn:P_diagonal2} and \eqref{eqn:P_nondiagonal2}. The parameters in $\mathbf{P}$ are $\phi= 0.5$, $\gamma= 0.95$, $\lambda=1$, $\psi= 1.5$ and $\mu = 1.0$.

We assume that the local patch foodwebs are linked by diffusive migration and therefore
$\mathbf{C}$ is given by,
\begin{align}
	\mathbf{C} = \begin{pmatrix}
		&10^{-4\cdot n_{1}}& 0 & 0&0 \\
		&0 &  10^{-4\cdot n_{2}} &0&0\\
        &0&0&10^{-4\cdot n_{3}} &0 \\
        &0&0&0&10^{-4\cdot n_{4}        } 
	\end{pmatrix} \;,
\end{align}
The entries in $\mathbf{C}$ scale with the inverse niche values and thus inverse bodymass,
and therefore larger species migrate slower than smaller species. This occurs for instance when small species can be dispersed passively by wind or water, while larger species require energy-consuming active dispersal, are more territorial, or overcome physical barriers less easily. 

The weighted laplacian $\mathbf{L}$ for the first spatial geometry used in Fig.~3a,d is given by
\begin{align}
	\mathbf{L} = c\cdot \begin{pmatrix}
		& 2 &-1 &-1 & 0 & 0 \\
		&-1 & 4 &-1 &-1 &-1 \\
        &-1 &-1 & 3 &-1 & 0 \\
        & 0 &-1 &-1 & 2 & 0 \\
        & 0 &-1 & 0 & 0 & 1 \\
	\end{pmatrix}
\end{align}
and the weighted laplacians used in Fig.~3b,e and Fig.~3c,f are given by
\begin{align}
	\mathbf{L} = c\cdot \begin{pmatrix}
		& 1 & 0 &-1 & 0 & 0 \\
		& 0 & 1 & 0 &-1 & 0 \\
        &-1 & 0 & 3 &-1 &-1 \\
        & 0 &-1 &-1 & 3 &-1 \\
        & 0 & 0 &-1 &-1 & 2 \\
	\end{pmatrix} \;.
\end{align}
For Fig.~3a-b, the global coupling strength $c$ is set to $0.045$, while it has a value of $0.09$ for Fig.~3c, which results in twice as large eigenvalues $\kappa$.
 
The eigenvalues $\lambda_\kappa$ of the full Jacobian are related to the eigenvalues $\kappa$ of $\mathbf{L}$ by the equation
\begin{equation}
	(\mathbf{P} - \kappa\mathbf{C}) \cdot \boldsymbol{q}
    = \lambda_{\kappa} \boldsymbol{q}\, .\label{qlambda2}
\end{equation}
The MSF is obtained by solving this equation for all $\kappa\in [0,\kappa_{max}]$ and plotting the real part of the leading eigenvalue $\text{Re} \left[ \lambda_{max}(\kappa)\right]$. In order to obtain the stability of a meta-foodweb for a given spatial topology, only the eigenvalues $\kappa$ of the respective Laplacian $\mathbf{L}$ are relevant.

\begin{table}
\begin{tabular}{|c|c|c|c|c|c|}
	\hline 
	$i$ & 1 & 2 & 3 & 4 \\ 
	\hline 
	$n_i$ & 0.97 & 0.34 & 0.91 & 0.12 \\ 
    	\hline
  	$c_i$ & 0.42 & 0.17 & 0.43 & 0.06 \\ 
    	\hline
   	$r_i$ & 0.74 & 0.34 & 0.49 & 0.11 \\ 
	\hline
\end{tabular}
\caption{Niche values, feeding ranges and centers used for the local foodweb in figure 2 rounded to the second decimal. }
\label{fig2:niche}
\end{table}
\FloatBarrier

\subsection{Details for Figure 4}

For Fig.~4, we generated a 20-species foodweb, with a connectivity, $C$, of $0.15\pm 0.01$.
The matrix $P$ for the foodweb was constructed using the parameters $\phi= 0.5$, $\gamma= 0.75$, $\lambda=1$, $\psi= 1.0$ and $\mu = 1.0$. 

The global coupling strength between the patches is set to $0.045$ for Fig.~4a and to $0.09$ for Fig.~4b. In addition to diffusive migration, the coupling matrix $\mathbf{C}$ contains cross-diffusion which describes adaptive migration, this is predators follow prey and prey species avoid their predator. Thus, for a predator-prey pair $(ij)$, the submatrix $\mathbf{C_{ij}}$ takes the form 
\begin{align}
	\mathbf{C_{ij}} = \begin{pmatrix}
		& 10^{-4\cdot n_{i}}& a \cdot10^{-4\cdot n_{i}} \\
		& - a \cdot10^{-4\cdot n_{j}} &  10^{-4\cdot n_{j}} 
	\end{pmatrix} \;.
\end{align}

with $a = 17.22$ for Fig.~4a and $a = 0.178$ for Fig~4b. All other non-diagonal entries in $\mathbf{C}$ are zero. The parameters used to create the 20-species foodwebs can be found in Tab.~\ref{tab:3a} and Tab.~\ref{tab:3b}.


\begin{table}
\begin{tabular}{|c|c|c|c|c|c|c|c|c|c|c|}
\hline
$i$ & 1 & 2 & 3 & 4 & 5 & 6 & 7 \\
\hline
$n_{i}$ & 0.02 & 0.69 & 0.12 & 0.20 & 0.54 & 0.81 & 0.58 \\
\hline
$r_{i}$ & 0.00 & 0.51 & 0.02 & 0.18 & 0.00 & 0.18 & 0.03 \\
\hline
$c_{i}$ & 0.010 & 0.275 & 0.060 & 0.100 & 0.490 & 0.650 & 0.115 \\
\hline
\hline
$i$ & 8 & 9 & 10 & 11 & 12 & 13 & 14 \\
\hline
$n_{i}$ & 0.65 & 0.56 & 0.54 & 0.51 & 0.74 & 0.81 & 0.80 \\
\hline
$r_{i}$ & 0.23 & 0.03 & 0.03 & 0.02 & 0.17 & 0.09 & 0.26 \\
\hline
$c_{i}$ & 0.525 & 0.245 & 0.235 & 0.420 & 0.555 & 0.495 & 0.340 \\
\hline
\hline
$i$ & 15 & 16 & 17 & 18 & 19 & 20 & \\
\hline
$n_{i}$ & 0.88 & 0.46 & 0.10 & 0.19 & 0.83 & 0.31 & \\
\hline
$r_{i}$ & 0.53 & 0.12 & 0.02 & 0.06 & 0.25 & 0.02 & \\
\hline
$c_{i}$ & 0.595 & 0.260 & 0.080 & 0.110 & 0.615 & 0.150 & \\
\hline
\end{tabular}
\caption{Niche values, feeding ranges and centers for the foodweb used in subgraph 3a.}
\label{tab:3a}
\end{table}


\begin{table}
\begin{tabular}{|c|c|c|c|c|c|c|c|c|c|c|}
\hline
$i$ & 1 & 2 & 3 & 4 & 5 & 6 & 7 \\
\hline
$n_{i}$ & 0.41 & 0.97 & 0.55 & 0.95 & 0.46 & 0.55 & 0.85 \\
\hline
$r_{i}$ & 0.23 & 0.49 & 0.22 & 0.75 & 0.20 & 0.27 & 0.01 \\
\hline
$c_{i}$ & 0.245 & 0.345 & 0.480 & 0.485 & 0.120 & 0.265 & 0.595 \\
\hline
\hline
$i$ & 8 & 9 & 10 & 11 & 12 & 13 & 14 \\
\hline
$n_{i}$ & 0.74 & 0.98 & 0.03 & 0.66 & 0.01 & 0.05 & 0.93 \\
\hline
$r_{i}$ & 0.16 & 0.18 & 0.02 & 0.06 & 0.00 & 0.02 & 0.43 \\
\hline
$c_{i}$ & 0.600 & 0.74 & 0.01 & 0.320 & 0.010 & 0.040 & 0.595 \\
\hline
\hline
$i$ & 15 & 16 & 17 & 18 & 19 & 20 & \\
\hline
$n_{i}$ & 0.53 & 0.38 & 0.49 & 0.70 & 0.76 & 0.93 & \\
\hline
$r_{i}$ & 0.41 & 0.03 & 0.30 & 0.07 & 0.12 & 0.24 & \\
\hline
$c_{i}$ & 0.305 & 0.035 & 0.260 & 0.455 & 0.550 & 0.120 & \\
\hline
\end{tabular}
\caption{Niche values, feeding ranges and centers for the foodweb used in subgraph 3b.}
\label{tab:3b}
\end{table}

\subsection{Details for Figure 5}

The spatial topologies were generated as random geometric graphs \cite{Dall2002}. The range $R$ is scaled with the number of nodes $N$ of the spatial graph
\begin{equation}
	R = \frac{r}{\sqrt N},
\end{equation}
where the parameter $r$ is the unscaled range. For the random geometric graph with $N=25$ used in Fig.~5a and 5c the parameter was set to $r=1.3$. The spatial graph with $N=500$ used in Fig.~5b and 5d was generated with $r=1.4$. Only connected graphs were selected.

\begin{figure}
	\includegraphics[scale=0.5]{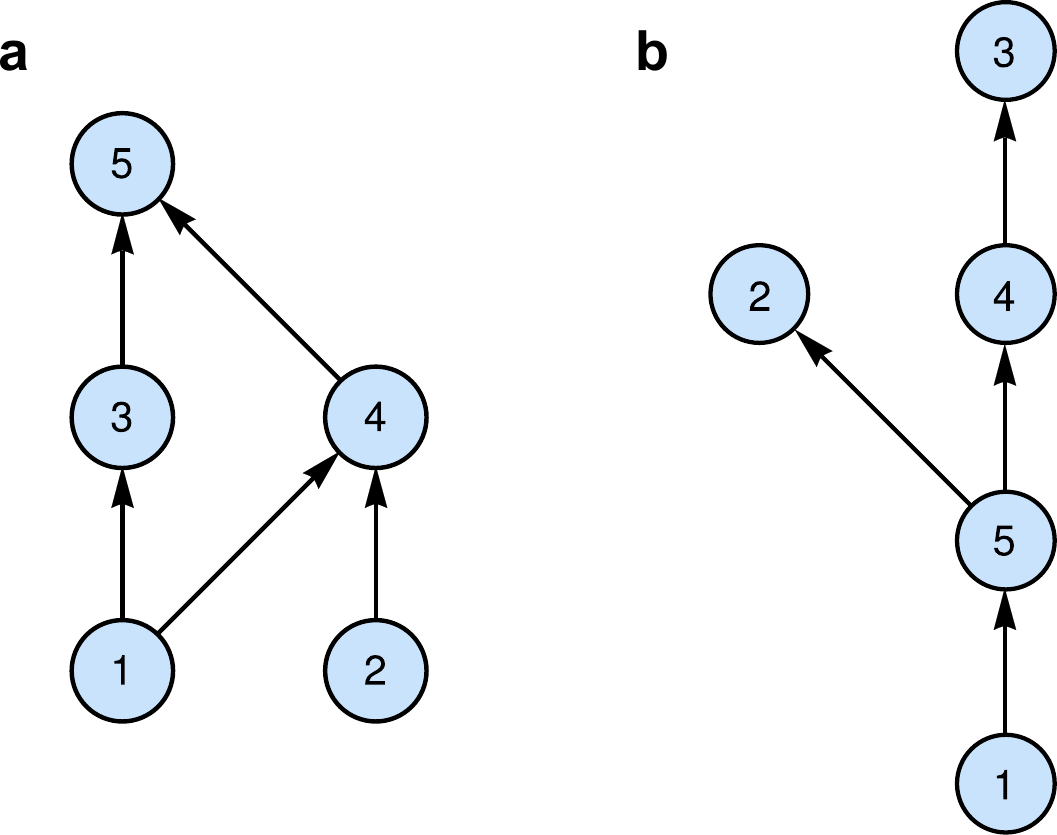}
    \caption{Five species foodweb, used in Fig.~4a and 4c (a) and for the animated oscillations (b). The shown numbers are the indices of the species. The arrows indicate the direction of biomass flow. In foodweb (a), species 1 and 2 are the primary producers and species 5 is the apex predator. In network (b) species 1 is the primary producer and species 3 is the observed top predator.}
    \label{fig:5species}
\end{figure}

For Fig.~5a and 5c, we used a manually constructed 5-species foodweb with the niche values shown in Table~\ref{fig4:niche} and a connectivity of $C=0.25$. The feeding links are defined by the weighted adjacency matrix of the local foodweb
\begin{equation}
	{\bf A} = \begin{pmatrix}
		0 & 0 & 0 & 0 & 0 \\
		0 & 0 & 0 & 0 & 0 \\
        1 & 0 & 0 & 0 & 0 \\
        \frac 1 2 & \frac 1 2 & 0 & 0 & 0 \\
        0 & 0 & \frac 1 2 & \frac 1 2 & 0 \\
	\end{pmatrix} \;.
\end{equation}
Species 1 and 2 are primary producers. Predators prey equally on the corresponding prey species. The resulting foodweb is shown in Fig.~\ref{fig:5species}a.

\begin{table}
\begin{tabular}{|c|c|c|c|c|c|}
	\hline 
	$i$ & 1 & 2 & 3 & 4 & 5 \\ 
	\hline 
	$n_i$ & 0.2 & 0.3 & 0.4 & 0.4 & 0.7 \\ 
	\hline
\end{tabular}
\caption{Niche values used for the local foodweb in figure 4 (a) and (c).}
\label{fig4:niche}
\end{table}

The population dynamics equations used for the explicit simulations are 

\begin{align}
\dot X_i^k
	&= G_i^k(X_i^k) - M_i^k(X_i^k) \notag \\
	&+ F_i^k(X_1^k, \ldots ,X_S^k) - \sum_j  \frac{R_{ji}^k(X_i^k) F_j^k(T_j^k,X_j^k)}{T_j^k (X_1^k,\ldots,X_S^k)} \notag \\
	&+ \sum_l \Big[ E_i^{kl}(X^k_1,\ldots,X^k_S,X^l_1,\ldots,X_S^l) \notag \\
	&\qquad\qquad - E_i^{lk}(X^l_1,\ldots,X^l_S,X^k_1,\ldots,X_S^k) \Big] \;,
\end{align}

with the functions

\begin{align}
	G_i^k(X_i^k) &= s_i X_i^k \label{eq:G} \;,\\ 
	M_i^k(X_i^k) &= p_i X_i^k + q_i {X_i^k}^2 \label{eq:M} \;,\\
	F_i^k(X_i^k, \ldots ,X_S^k) &= \frac{\sum_j a_i A_{ij} X_i^k X_j^k}{1+\sum_j a_i h_i A_{ij} X_j^k} \notag \\
	&= F_i^k(T_i^k, X_i^k) \notag \\
	&= \frac{a_i T_i^k X_i^k}{1+ a_i h_i T_i^k} \label{eq:F} \;,\\
	R_{ij}^k &= A_{ij} X_j^k \label{eq:C} \;,\\
	T_i^k(X_1^k, \ldots , X_S^k) &= \sum_j A_{ij} X_j^k \label{eq:T} \;,\\
	E_i^{kl}(X_i^l) &= c_i^{kl} X_i^{l} \;. \label{eq:E}
\end{align}

The predation $F_i^k(X_i^k)$ is based upon the Holling-Type II functional response. Migration $E_i^{kl}(X_i^k)$ and primary production $G_i^k(X_i^k)$ are in linear proportion to the population sizes while respiration and mortality $M_i^k$ are between linear and quadratic. $R^k_{ij}$ is the relative contribution of species $j$ to the prey consumed by $i$. $T_i^k$ is the total amount of food available to species $i$. The adjaceny matrix, $A_{ij}$, contains the information about the feeding links. The migration term $E_i^{kl}$ scales linearly with population size, i.e.~migration is diffusive. The used parameters are shown in Tab.~\ref{tab:ExplicitParameters}, and the body mass was  calculated as
\begin{equation}
	m_i = 10^{3n_i} \;.
\end{equation}

\begin{table}
\begin{tabular}{c|c|c}
Parameter & Meaning & Value\\
\hline
$a_i$ & Attack rate of Pred. & 4$m_{i}^{-0.25}$\\
$h_i$ & Handling time of Pred. & 0.26$m_{i}^{-0.25}$\\
$p_i$ & Linear Mort. Coeff. & 0.52$m_{i}^{-0.25}$\\
$q_i$ & Quadratic Mort. Coeff. & 0.34$m_{i}^{-0.25}$\\
$c_i$ & Diffusion Coeff. & $10^{-3}$ $m_{i}^{0.75}$\\
$s_i$ & Primary Prod. Coeff. &  4.5$m_{i}^{-0.25}$\\
\end{tabular}
\caption{Parameters for the explicit model in Eqs.~(\protect\ref{eq:G}) to (\protect\ref{eq:E}), and the values used when calculating explicit population dynamics. The parameter $s_i$  is nonzero only for primary producers. \label{tab:ExplicitParameters}}
\end{table}

If a real leading eigenvalue changes its sign from negative to positive, the system undergoes a Turing instability (Fig.~5). If the leading eigenvalue is complex, a wave instability occurs, which leads to (at least transient) spatio-temporal oscillations. An animation of such an oscillating system is available at \url{http://eco.fkp.physik.tu-darmstadt.de/drossel/gramlich/animation.gif}.

In order to investigate the dynamics after an instability occurred, trajectories starting close to the unstable homogeneous state were simulated until they approached a new long term behavior. For these simulations initial values were chosen randomly with a maximum relative distance of 0.1\% to the homogeneous steady state.

\FloatBarrier

\begin{table}
\begin{tabular}{|c|c|c|c|c|c|}
	\hline 
	$i$ & 1 & 2 & 3 & 4 & 5 \\ 
	\hline 
	$n_i$ & 0.042 & 0.865 & 0.990 & 0.614 & 0.257 \\ 
	\hline
    ${\alpha_P}_i$ & 0.824 & 0.019 & 0.010 & 0.0592 & 0.306 \\
    \hline
    $\delta_i$ & 0 & 1 & 1 & 1 & 1 \\
    \hline
    $\sigma_i$ & 0.512 & 0.000 & 0.000 & 0.495 & 0.668 \\
    \hline
    \hline
    $i,j$ & $2,5$ & $3,4$ & $4,5$ & $5,1$ & else\\
    \hline
    $\beta_{ij}$ & 0.457 & 1.000 & 0.543 & 1.000 & 0.000 \\
    \hline
    $A_{ij}$ & 1 & 1 & 1 & 1 & 0 \\
    \hline
\end{tabular}
\caption{Parameters used for the local foodweb in the animation.}
\label{tab:animation_niche}
\end{table}

\subsection{Details for the animation}
The animation shows a comparison of the respective eigenvector (open circles) and the observed deviation from the homogeneous value in the final state (dots). In the right panel, the population density of the top predator (index 3, Fig.~\ref{fig:5species}b) is shown. The homogeneous steady state is located at a biomass of $1$, larger biomasses are shown in orange and smaller values in blue.
\par
The foodweb of the oscillating system (Fig.~\ref{fig:5species}b) was generated with the niche model. The explicit model was matched to the parameters $\phi= 0.5$, $\gamma= 0.75$, $\psi= 1.5$, and $\mu = 1.0$. Species depended parameters can be found in Tab.~\ref{tab:animation_niche}, and the diffusion coefficient is given by,
\begin{align}
	c_i = 10\cdot 10^{-8n_i} \;.
\end{align}
The population dynamic is given by
\begin{align}
	\dot x_i^k = {\alpha_P}_i^k
	\Bigg[
		&\quad \; \tilde \delta_i^k g_i^k(x_i^k) 
		+ \delta_i^k f_i^k(t_i^k,x_i^k) \notag\\
		&- \tilde \sigma_i^k m_i^k(x_i^k)
		- \sigma_i^k \sum_j \beta_{ji}^k d_{ji}^k(x_1^k,\ldots,x_S^k) \Bigg]\notag\\
		&- \sum_l L^{kl} c_i x_i^{l} \;,
\end{align}
with
\begin{align}
	g_i^k(X_i^k) &= \left(x_i^k\right)^\phi  \;,\\ 
	m_i^k(X_i^k) &= \left(x_i^k\right)^\mu \;,\\
	f_i^k(X_i^k, \ldots ,x_S^k) &= t_i^k \left(x_i^k\right)^\psi \frac{1+K}{t_i^k+K}\;,\\
	d_{ij}^k &= \left(x_i^k\right)^\psi x_j^k \frac{1+K}{T_i^k+K} \;,\\
	t_i^k(x_1^k, \ldots , x_S^k) &= \sum_j A_{ij} x_j^k \;,
\end{align}
where
\begin{equation}
	K = \frac{\gamma}{1-\gamma} \;.
\end{equation}

\end{document}